\documentclass[sigconf]{acmart} 
\usepackage{float}
\AtBeginDocument{%
  \providecommand\BibTeX{{%
    \normalfont B\kern-0.5em{\scshape i\kern-0.25em b}\kern-0.8em\TeX}}}
\usepackage{bbding}
\usepackage{graphicx}
\usepackage{amsmath}
\usepackage{makecell}
\usepackage{float}
\usepackage{color}
\usepackage{booktabs}
\usepackage{subfig}
\usepackage{hyperref}
\usepackage{xcolor}
\usepackage{multirow} 
\usepackage{pifont}
\setcopyright{acmlicensed}
\copyrightyear{2026}
\acmYear{2026}
\setcopyright{cc}
\setcctype{by}
\acmConference[CHI '26]{Proceedings of the 2026 CHI Conference on Human Factors in Computing Systems}{April 13--17, 2026}{Barcelona, Spain}
\acmBooktitle{Proceedings of the 2026 CHI Conference on Human Factors in Computing Systems (CHI '26), April 13--17, 2026, Barcelona, Spain}
\acmPrice{}
\acmDOI{10.1145/3772318.3791238}
\acmISBN{979-8-4007-2278-3/2026/04}

\definecolor{ccolor}{RGB}{55, 120, 66}

\begin{document}

\author{Zhihan Jiang}
\email{zj2445@cumc.columbia.edu}
\orcid{0000-0003-4857-7143}
\affiliation{
\institution{The University of Hong Kong}
\city{Hong Kong, SAR}
\country{China}\\
\institution{Columbia University}
\city{New York}
\country{United States}
}

\author{Qianhui Chen}
\email{2023103597@ruc.edu.cn}
\orcid{0009-0000-5526-284X}
\affiliation{
\institution{Renmin University of China}
\city{Beijing}
\country{China}
}

\author{Chu Zhang}
\email{zhangchu0908@outlook.com}
\orcid{0009-0004-4491-2279}
\affiliation{
\institution{City University of Hong Kong\\Studio for Narrative Spaces}
\city{Hong Kong, SAR}
\country{China}\\
\institution{Guangdong University of Technology}
\city{Guangzhou}
\country{China}
}

\author{Yanheng Li}
\email{lydia.yh-li@my.cityu.edu.hk}
\orcid{0000-0002-9767-3468}
\affiliation{
\institution{City University of Hong Kong\\Studio for Narrative Spaces}
\city{Hong Kong, SAR}
\country{China}
}

\author{RAY LC}
\authornote{Correspondences can be addressed to ray.lc@cityu.edu.hk.}
\email{ray.lc@cityu.edu.hk}
\orcid{0000-0001-7310-8790}
\affiliation{
\institution{City University of Hong Kong\\Studio for Narrative Spaces}
\city{Hong Kong, SAR}
\country{China}}

\title[Hear You in Silence]{\emph{Hear You in Silence}:\\Designing for Active Listening in Human Interaction with Conversational Agents Using Context-Aware Pacing}

\begin{teaserfigure}
    \centering
    \includegraphics[width=.82\linewidth]{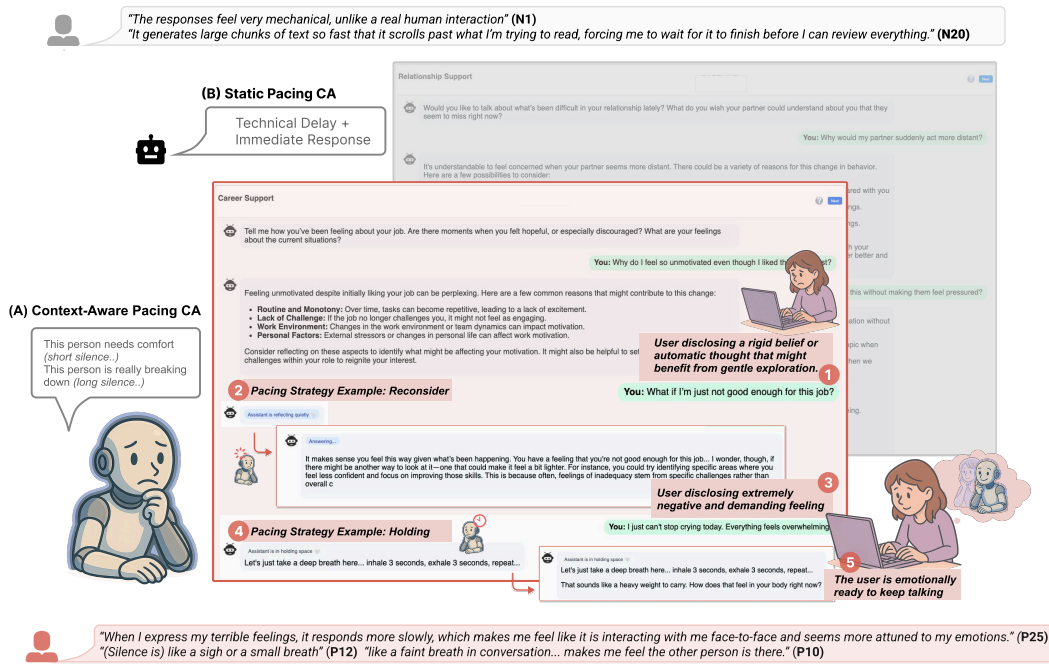}
    \caption{Comparison of two conversational pacing models. \textbf{(A)} A Conversational Agent (CA) with Context-Aware Pacing adapts its response timing to the situation. For example, when a user shares a rigid belief \ding{172}, the agent employs a \texttt{Reconsider} strategy \ding{173}, briefly pausing to reflect before responding. When encountering intense negative emotions \ding{174}, it uses a \texttt{Holding} strategy \ding{175}, creating a prolonged, supportive silence that allows the user to become emotionally ready to continue. \textbf{(B)} In contrast, a CA with Static Pacing defaults to technical delays and immediate responses regardless of context.}
    \label{fig:teaser}
    \Description{Comparison of conversational agent pacing strategies: context-aware adaptive pacing versus vs. pacing with fixed delays. The figure shows two pacing models. The lower part (A) depicts a Conversational Agent with Context-Aware Pacing, where the response timing dynamically adapts to different conversational contexts (e.g., info-seeking or emotional disclosure). Different types of input trigger distinct pacing strategies. The upper part (B) illustrates a CA with Static Pacing that employs a uniform timing approach, such as immediate responses or fixed technical delays, regardless of the input context. For the context-aware pacing CA, there are two pacing examples. When a user shares a rigid belief or automatic thought that could benefit from gentle exploration, the CA responds with the Reconsider pacing strategy, pausing to reflect before responding. When a user expresses extremely negative or demanding emotions, the CA utilizes the Holding pacing strategy, creating a supportive silence that allows the user to become emotionally ready to continue.}
\end{teaserfigure}

\begin{abstract}
In human conversation, empathic dialogue requires nuanced temporal cues indicating whether the conversational partner is paying attention. This type of "active listening" is overlooked in the design of Conversational Agents (CAs), which use the same pacing for one conversation. To model the temporal cues in human conversation, we need CAs that dynamically adjust response pacing according to user input. We qualitatively analyzed ten cases of active listening to distill five context-aware pacing strategies: \emph{Reflective Silence}, \emph{Facilitative Silence}, \emph{Empathic Silence}, \emph{Holding Space}, and \emph{Immediate Response}. In a between-subjects study (N=50) with two conversational scenarios (relationship and career-support), the context-aware agent scored higher than static-pacing control on perceived human-likeness, smoothness, and interactivity, supporting deeper self-disclosure and higher engagement. In the career-support scenario, the CA yielded higher perceived listening quality and affective trust. This work\footnote{Project website: \url{https://zhihanjiang.com/pacing/}.} shows how insights from human conversation like context-aware pacing can empower the design of more empathic human-AI communication.
\end{abstract}




\maketitle

\section{Introduction}\label{sec:Introduction}
Human connection relies on more than just words; it is profoundly shaped by the subtle, temporal rhythms of communication \cite{bruneau1973communicative}. In supportive conversations, ``active listening'' requires the listener to use nuanced pacing to signal empathy, attention, and shared understanding \cite{weger2014relative}. Silence, in this context, is not an empty void but a powerful communicative tool to show attention and interest \cite{bruneau1973communicative}. It can be used strategically to hold space for a speaker, encourage deeper self-disclosure, or convey empathy \cite{levitt2001sounds, back2009compassionate}. Understanding the function of this pacing is fundamental to understanding how people build trust and rapport.

Despite the increasing integration of Conversational Agents (CAs) into socially significant roles like emotional support, their design largely ignores subtle but crucial forms of human communication such as pacing \cite{gnewuch2022opposing}. Current systems typically adopt a static and mechanistic interaction style, prioritizing efficiency and immediate responses \cite{10.1145/3706598.3713503, nandy2025balancing}. This efficiency-first paradigm creates a critical disconnect, resulting in interactions that feel robotic and superficial \cite{gnewuch2022opposing}. The core problem is not just technological, but a failure to model the interpersonal functions of conversational pacing that are essential for human-centric communication \cite{templeton2023listening, templeton2022fast}. However, historically in the field of CA design, dynamic pacing with respect to textual content has been largely ignored.

Previous research attempting to bridge this gap has faced two major limitations, stemming from an incomplete understanding of the dynamic pacing in human communication:
\begin{itemize}
    \item \textbf{Superficial Models of Timing:} Attempts to manage conversation timing in CAs have often treated pauses as simple, static delays—technical signals to simulate ``thinking'' or reduce perceived automation \cite{park2019slow, 10.1145/3640794.3665550}. This perspective overlooks the potential for silence to serve as a dynamic, relational tool embedded within the conversation's context.
    \item \textbf{Over-emphasis on Verbal Content:} Research on implementing active listening in CAs has focused almost exclusively on verbal strategies, such as paraphrasing and asking follow-up questions \cite{collins2022listening}. While valuable, these efforts ignore the non-verbal dimension of pacing, which communication psychology has long established as an intentional communicative act essential for building rapport \cite{back2009compassionate}.
\end{itemize}

What existing works fundamentally overlook is that human communication is highly context-dependent \cite{templeton2023listening}. The appropriateness of a particular pacing strategy shifts based on relational goals \cite{lim2025proactivity}. For instance, while immediate responses suit task-oriented goals, supportive contexts demand more patient and reflective pacing to foster connection \cite{kang2018study}. Therefore, to design more human-centric and supportive CAs, there is a need to deepen our understanding of how people use pacing strategies in different social contexts.

To address this gap, our study first seeks to understand and systematize human pacing strategies before translating them into computational design. We conducted a qualitative analysis of ten exemplary video recordings of human-to-human active listening to distill the functional uses of silence and response timing. This analysis yielded a taxonomy of five context-aware pacing strategies: \emph{Reflective Silence}, \emph{Facilitative Silence}, \emph{Empathic Silence}, \emph{Holding Space}, and \emph{Immediate Response}, corresponding to eight specific pacing strategies (\texttt{Recognize}, \texttt{Reconfirm}, \texttt{Re-engage}, \texttt{Reposition}, \texttt{Reconsider}, \texttt{Resonate}, \texttt{Holding}, and \texttt{Resolve}). Based on these findings, we designed and built an LLM-based CA \cite{zeng_ronaldos_2025,zhou_eternagram_2024} that uses context-aware pacing embodying these strategies. 
With this CA, our work investigates the value of operationalizing the relational functions of human silence into a mechanistic, implementable framework of context-aware pacing. We hypothesized that users would perceive this pacing as empathetic social cues, facilitating more empathic human-AI communication.
The CA dynamically adjusts the response pacing strategies according to user input. More specifically, this work explores these two research questions:

\begin{itemize}
    \item \textbf{RQ1:} How do CAs using context-aware pacing strategies impact users' perceived quality of interaction and experience, specifically in terms of listening quality, affective trust, cognitive trust, human-likeness, smoothness, and interactivity?
    \item \textbf{RQ2:} How does context-aware pacing influence user interaction behaviors in text-based supportive conversations with CAs, specifically in terms of depth of self-disclosure and level of engagement?
\end{itemize}

To answer these questions, we conducted a between-subject study (N=50) comparing our context-aware CA against a static-pacing baseline CA across two supportive scenarios (career and relationship advice). 
These contexts were designed to contrast a scenario blending practical advice with emotional needs (career support) against one centered on interpersonal and affective processing (relationship support). Our work makes the following contributions:

\begin{itemize}
    \item We challenge the prevailing focus on verbal output in CA design. To the best of our knowledge, we are the first to systematically introduce conversational pacing into CA design, providing empirical evidence that the strategic use of silence for context-aware pacing is critical for developing empathic human-AI relationships analogous to active listening.
    \item We designed and evaluated a context-aware pacing CA under two supportive scenarios (career and relationship). Our study systematically investigates the effects of context-aware pacing on perceived interaction quality and experience (i.e., listening quality, affective trust, cognitive trust, human-likeness, smoothness, and interactivity) and interaction behaviors (i.e., depth of self-disclosure and level of engagement).
    \item We propose a new interaction design space centered on dynamic pacing modulation and introduce five concrete, implementable mechanisms: \emph{Reflective Silence}, \emph{Facilitative Silence}, \emph{Empathic Silence}, \emph{Holding Space}, and \emph{Immediate Response}. Our work provides practical pathways and design implications for more empathic human-AI communication.
\end{itemize}

\section{Related Work}\label{sec:Background}
\subsection{Active Listening in Conversational Agents}
\label{sec:related_work_active}
Active listening is a dynamic process that combines attentiveness, understanding, and constructive intention \cite{kluger2022power}. It conveys unconditional acceptance and unbiased attitudes towards speakers' experiences \cite{rogers1959theory}. Rooted in the principle of empathic listening, psychotherapists have developed several active listening strategies, such as eye contact, head-nodding, backchanneling, paraphrasing, and summarizing \cite{weger2010active, 10.1145/3706598.3714228}. Despite being subtle, these cues for listening could produce positive interaction outcomes \cite{weger2014relative}, such as reducing loneliness \cite{itzchakov2023connection}, improving trustworthiness \cite{ramsey1997listening}, responsiveness \cite{itzchakov2022foster}, and friendliness \cite{bodie2012listening}.

One of the primary benefits of active listening is its power to encourage deeper self-disclosure. Guiding individuals to express themselves more deeply has been a central topic in HCI studies \cite{zhang2023tools}. In human-to-human communication, people tend to share more about themselves when they perceive clear benefits or feel emotionally safe \cite{hallam2017online, 10.1145/3613904.3642385,derlega2008does}. One key barrier to deeper disclosure lies in the anxiety of social evaluation, which refers to the distress or fear experienced in situations where an individual anticipates being judged by others \cite{derlaga1987self}. In contrast, active listening demonstrates a non-judgmental and open attitude towards speakers, thereby encouraging more expression. 

Inspired by these benefits, HCI researchers have explored integrating active listening capabilities into CAs. 
In voice-based CAs, backchannel cues (\textit{``hmm,''} \textit{``yeah''}) could significantly promote a user's sense of being ``heard'' \cite{10.1145/3555164, 10.1145/3313831.3376131, 10.1145/3708557.3716343}. Similarly, verbal strategies, such as paraphrasing, verbalizing emotions, summarizing, and encouraging disclosure, also have positive effects in conveying attentiveness and improving user experience with CAs \cite{10.1145/3313831.3376131}. Social Penetration Theory suggests that the exchange of personal information catalyzes relationship development through reciprocity \cite{altman1973social}. Similar to interpersonal communication \cite{nass2000machines}, text-based CAs that disclose their feelings, emotions or preferences could create an equal and affable conversational atmosphere \cite{meng2021emotional}, thereby eliciting a sense of being ``heard'' and enhancing people's perceived intimacy \cite{10.1145/3313831.3376175}. More recently, researchers have also proposed displaying the agent's ``internal empathic resonance'' via a secondary text channel to explicitly visualize its understanding of the user \cite{schmidmaier2025using}.

Despite this progress, existing studies primarily focus on \textit{what} an agent says, overlooking a fundamental component of human communication: \textit{how} and \textit{when} it responds. In human dialogue, the non-verbal, temporal dynamics of a conversation (pacing) are powerful communicative signals that convey empathy and presence \cite{bruneau1973communicative, weger2014relative}. By treating active listening primarily as a content-generation problem, existing CA designs have neglected the crucial role of conversational pacing. Our work addresses this critical gap by investigating how strategic silence and timing can be designed to foster deeper connection in human-AI interaction.

\subsection{Conversational Pacing for Active Listening}
\label{sec:related_work_pacing}
In human conversation, appropriate conversational pacing is a key non-verbal component of active listening. The optimal pace is context-dependent: a fast response could signal proactivity in simple requests \cite{templeton2023listening, templeton2022fast}, whereas slower pacing could provide more space for people to reflect and process complex issues, enacting feelings of being understood and connected \cite{bartels2016eloquent}. In slower pacing mode, silence is used strategically to convey rich emotional and informational meanings \cite{templeton2023listening}. Unlike unconscious and simple linguistic pauses within conversations, interactive silence is a deliberate tool used to convey intimacy and deep thought \cite{bruneau1973communicative}. In this context, silence contributes to creating a space for softening people's fierce emotions, leading to shared understanding and nonjudgmental attitudes \cite{hill2003therapist}. For example, silence is especially emphasized in patient-clinician encounters \cite{levitt2001sounds, knutson2015varieties}. In end-of-life care, invitational and compassionate silence given by doctors could be consciously trained through contemplative practice, contributing not only to empathy and even ``mutual wisdom'' towards life \cite{back2009compassionate}. While conversational pacing is a critical component of active listening in human communication, its systematic design and evaluation within CAs remains underexplored \cite{kafaee2024silence}.

High-quality listening is not associated with a uniformly fast or slow pace. An inappropriately fast reply may feel mechanical and shallow \cite{gnewuch2022opposing}, while a persistently slow response can cause confusion, reduce naturalness, and irritate the user \cite{levitt2001sounds, 10.1145/3472307.3484181, kum2022can}. 
However, within this spectrum, deliberate pacing is often more than just delays in response time. Slower pacing can potentially enhance social presence \cite{holtgraves2007perceiving, holtgraves2007procedure, gnewuch2022opposing}. This perspective closely aligns with the CASA (Computer As Social Actor) framework, which suggests that people will mindlessly perceive machines demonstrating enough social cues as if they were human \cite{reeves1996media, nass2000machines}. Even minimal social cues (such as gender or ethnicity labels) could trigger unconscious social responses to machines, regardless of the specific demographic or background of the users \cite{nass2000machines}. Under this framework, silence, as a fundamental element in interpersonal communication, is not merely a delay in response time but a communicative signal. Similar to human communication, slower pacing could lead to a more natural and smooth experience \cite{appel2012does} and trigger the effort heuristic \cite{10.1145/3449287, kruger2004effort}, thereby enhancing social presence perceptions \cite{gnewuch2022opposing}.
However, slower pacing is also frequently associated with unresponsiveness, especially in situations where a fast response is expected \cite{gnewuch2022opposing}. 
This trade-off highlights the need for an adaptive approach to pacing \cite{hauptman2025ethical}. The effectiveness of a CA's pacing is highly dependent on the context, as users have different expectations for task-oriented conversations (prioritizing efficiency) versus socio-emotional conversations (prioritizing support) \cite{lim2025proactivity}. 

Building on this, we introduce context-aware pacing: a strategy where a CA flexibly adjusts its response timing to a user's situational needs on a turn-by-turn basis. Instead of adopting a static pace, such a context-aware pacing CA could use strategic silence to better convey attentiveness and empathy.

\subsection{The Role of Delay in Human-AI Interaction}
\label{sec:related_work_artifact}
Existing works relevant to the temporal aspects of CAs mainly focus on investigating the technical delay before AI responses, implicitly treating the response delay as a technical flaw, i.e., a system artifact to be mitigated or concealed \cite{10.1145/3719160.3736636}. 
By taking on this position, various strategies were explored to reduce negative user perceptions of delayed responses. One common strategy is using textual or graphic indicators to show progress. Some interactive indicators aim at actively involving users in other activities during waiting, shifting users' attention away from slow pacing, and decreasing their perceived waiting time \cite{10.1145/3552327.3552329}. More recent studies used human-like behaviors such as backchannel fillers to mitigate latency during interaction, which would enhance AI's interactivity \cite{lopez2019testing, 10.1145/3290607.3312913, 10.1145/3472307.3484181, christenfeld1995does, 7745134}. Current LLM-based conversation tools also use some visual cues to indicate progress. For example, ChatGPT\footnote{https://chatgpt.com/} uses a pulsating black dot to indicate ongoing processing, displaying messages in real time as they are generated ``with great effort''. Beyond traditional progress indicators, some studies also call for a more ``meaningful'' design to demonstrate the uniqueness of a Generative AI system, such as presenting AI's reasoning steps \cite{10.1145/3706599.3719725} (e.g., Gemini-2.5-pro\footnote{https://gemini.google.com/} and GPT-5 Thinking).

Considering the potential trade-offs of slow pacing design, these strategies inform when exploring the communicative potential of machine's ``silence'', how to minimize its downsides on user experience. However, this perspective overlooks evidence that a slower response is not always detrimental \cite{yang_ai_2022}. For example, Park et al. \cite{park2019slow} found that when an algorithm provided accurate advice, slower response times actually led to increased user trust and adherence to its suggestions. This finding suggests that ``silence'' is not inherently negative; its interpretation is highly context-dependent \cite{10.1145/3472307.3484181}. Moreover, as discussed in Sections \ref{sec:related_work_active} and \ref{sec:related_work_pacing}, existing studies overlooked the temporal cues to convey communicative and socio-emotional meaning \cite{templeton2023listening}. Silence can be perceived not just as a system delay but as thoughtful deliberation. To bridge this gap, we designed and evaluated a context-aware pacing CA to provide empirical evidence on how context-aware pacing influences users' perceived interaction quality, experience, and interaction behaviors.

\section{Design Space}\label{sec:Designing}
\subsection{Formative Study on Context-Aware Pacing for Active Listening}

Most research on active listening in CAs focuses on verbal strategies like paraphrasing and questioning \cite{10.1145/3313831.3376131, Tustonja2024active}, leaving the crucial role of pacing within conversations largely unaddressed. This overlooks how silence functions as a powerful communicative resource in human interaction for emotional articulation and self-awareness \cite{gutierrez2024exploration}. We reframe silence as context-aware pacing, i.e., the strategic use of silence to support a speaker's needs. To understand how to operationalize this concept, we conducted a formative study analyzing real-world active listening cases, focusing specifically on classifying the function of pacing strategies in response to the conversational context.

\subsubsection{Active Listening Case Collection}
To ensure the validity and diversity of cases, we sourced videos from YouTube\footnote{https://www.youtube.com/} using the term ``Active Listening Counseling.'' We selected videos that explicitly mentioned ``active listening'' in their title or description. The videos were further screened based on the following criteria: (1) content centered around real or simulated counseling dialogues; (2) duration of over 20 minutes, including complete conversational segments; (3) clear visibility of listener behavior, including both verbal and nonverbal strategies; and (4) English as the primary language. From this corpus, we selected 10 cases (ranging in length from 20 to 60 minutes each) covering diverse topics (e.g., exam anxiety, relationship issues, body image) to ensure the generalizability of our findings. The links to the 10 cases can be found in the Appendix \ref{appx:links}.

\begin{figure*}[t]
\centering
\includegraphics[width=\textwidth]{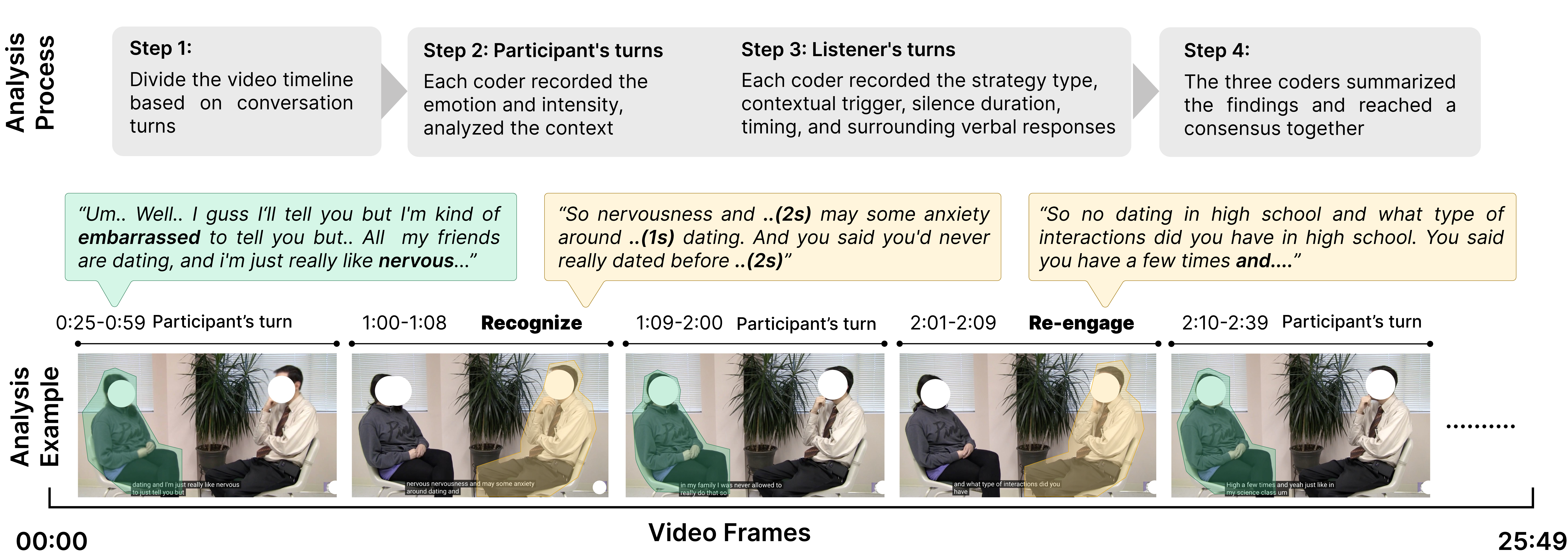}
\caption[example]{Data analysis process and an analysis example for a video where the speaker faced dating anxiety\footnotemark.}
\label{fig:timestamp}
\Description{This figure illustrates the qualitative data analysis methodology, divided into a process overview and a specific application example. The top section outlines a four-step linear workflow: dividing the video timeline by conversation turns, coding participant turns for emotion and intensity, coding listener turns for pacing strategies (including triggers and silence duration), and establishing consensus among coders. The bottom section visualizes this process applied to a video timeline (00:00–25:49) where a speaker discusses dating anxiety; it maps specific timestamps to video frames and transcripts, demonstrating how the listener employs the "Recognize" strategy (validating nervousness with pauses) and the "Re-engage" strategy (using prompts to continue the narrative) in response to the participant's disclosures.}
\end{figure*}
\footnotetext{Dating Anxiety: https://www.youtube.com/watch?v=Xj3q96mCfC8}
\subsubsection{Data Analysis}
We employed an open coding approach to analyze the function of pacing within active listening strategies employed by the counselors \cite{miles2014Quali}. Three researchers independently coded the data, focusing on ``strategic use of silence segments'' as the unit of analysis. Rather than treating silence as an isolated phenomenon, we analyzed how it was embedded in context, responded to prior utterances, and guided the dialogue. As shown in Figure~\ref{fig:timestamp}, our coding scheme captured dimensions such as the strategy type, contextual trigger, silence duration, timing, and surrounding verbal responses. Any disagreements were resolved through discussion to reach a consensus.

During the coding process, we reached saturation, meaning that no new categories emerged as we continued analyzing the data. The categories identified effectively covered all instances of silence and its strategic functions, indicating that we had fully captured the relevant behaviors within this context. To ensure consistency and reliability in the coding, we adopted a multi-coder approach. After each researcher independently coded the data, we compared their results and resolved any discrepancies through discussion. This discussion process helped us maintain consistency while ensuring the accuracy of the coding scheme.

\subsubsection{Pacing-Based Listening Strategies}
Ultimately, we identified five key types of pacing-based strategies, each serving distinct communicative and psychological support functions within conversations: \emph{Reflective Silence}, \emph{Facilitative Silence}, \emph{Empathic Silence}, \emph{Holding Space}, and \emph{Immediate Response}. \emph{Reflective Silence} conveys the listener's thoughtful processing and confirmation of the speaker's message. Through brief silence and reflective responses, the listener signals that they are seriously attending to the content, thereby reinforcing trust and engagement. \emph{Facilitative Silence} is employed in contexts of ambiguity, disruption, or topic transition. Here, silence acts as a subtle prompt, encouraging the speaker to clarify thoughts or re-engage in narration. \emph{Empathic Silence} broadly applies when speakers articulate deep feelings, beliefs, or values. By remaining silent, the listener expresses acceptance and empathy without judgment, fostering self-expression and emotional integration. \emph{Holding Space} emphasizes creating a non-directive, emotionally safe environment during moments of intense vulnerability. The listener uses silence to allow the speaker to fully experience and process their internal state without interruption or pressure. Finally, \emph{Immediate Response}, which applies no deliberate silence, addresses explicit information-seeking needs efficiently.
These strategies represent distinct modes of support in active listening that help respond to emotional disclosures, clarify narratives, and facilitate deeper conversational progression.
More specifically, they are operationalized by eight specific, implementable strategies (e.g., \texttt{Recognize}, \texttt{Reconfirm}, etc.), which are detailed in Table~\ref{tab:silence_strategies}.

\begin{table*}[t]
\small
\caption{Summary of Context-Aware Pacing Strategies in Active Listening}
\label{tab:silence_strategies}
\resizebox{\textwidth}{!}{%
\begin{tabular}{p{1.1cm} p{1.3cm} p{4.5cm} p{3.5cm} p{3.5cm} p{4.2cm} p{1.3cm} p{2cm} p{1.3cm}}
\toprule
\textbf{Type} & \textbf{Strategy} & \textbf{Context Trigger} & \textbf{Context Example} & \textbf{Description of Strategy} & \textbf{Strategy Example} & \textbf{Silence Duration} & \textbf{Timing} &  \textbf{Frequency} \\
\midrule

\emph{Reflective Silence}
& \texttt{Recognize} 
& When users need their experiences or feelings to be acknowledged or validated, or when they are seeking advice. 
& People: \textit{``I just don't know if I can keep going with...''}
& Provide emotional validation and reflect it back to them, with the goal of making the user feel understood.
& Listener: \textit{``OK, I see. Maybe... \underline{(silence)} You might be because...\underline{(silence)} Is that your feelings, right?''}
& 1–2s 
& After transition words like \textit{``but''}, \textit{``maybe''}, etc. & 21.5\% \\
\midrule

\emph{Facilitative Silence}
& \texttt{Reconfirm} 
& When the user says something vague, contradictory, or unclear, prompting you to reconfirm. 
& People: \textit{``I can't quite explain it. It's just... complicated.''}
& Rephrase the key information shared by the user in a gentle manner and invite them to expand on their narrative.  
& Listener: \textit{``\underline{(silence)} Can you tell me a bit more about what you mean?''}
& 2–3s 
& Before response
 & 27.3\%\\

& \texttt{Re-engage} 
& The user's story fades out, they pause awkwardly, or stop elaborating, needing a gentle nudge to continue. 
& People: \textit{``Hmm... \underline{(long silence)}''}
& Offer short, incomplete connecting phrases to encourage the user to continue adding to their story, ending with ellipses to maintain a sense of pause. 
& Listener: \textit{``So... And because...''}
& 2-3s 
& Before response
& 4.2\%\\
\midrule

\emph{Empathic Silence}
& \texttt{Reposition} 
& The user seems stuck in a rigid or negative perspective, expressing blame or hopelessness. 
& People: \textit{``I'll never get anywhere. I'm always stuck.''}
& After reflecting the user's emotions, guide them to re-examine their feelings or situation, offering positive exploratory suggestions.
& Listener: \textit{``I hear you feel stuck, \underline{(silence)} but have you thought about any small changes you could try?''}
& 5–6s 
& Before response
& 4.2\%\\

& \texttt{Reconsider} 
& The user expresses a rigid belief or automatic thought that might benefit from gentle exploration. 
& People: \textit{``People like me just don't succeed.''}
& Listener pauses to consider if the response may hurt. 
& Listener: \textit{``\underline{(silence)} It's difficult to face the (situations/problems). How did that experience make you feel?''}
& 2–3s 
& Before response
& 5.9\%\\

& \texttt{Resonate} 
& Whether positive or negative, the user is still immersed in the emotion of their story 
& People: \textit{``I had a... I was going through a lot... like I'm gonna be free for the rest of my life.''}
& Whether positive or negative, the listener should first pause, allowing the emotional flow to occur, and then respond emotionally.
& Listener: \textit{``\underline{(silence)} I feel quite touched in a particular kind of way.''}
& 3–15s 
& Before response
& 5.9\%\\
\midrule

\emph{Holding Space}
& \texttt{Holding}  
& The user repeatedly shares something intense, painful, or vulnerable. 
& People: \textit{``It still feels like it's all happening, like a nightmare I can't wake up from, so overwhelming.''}
& When the user shares deeply negative emotions or pain, appropriately guide emotional release.
& Listener: \textit{``\underline{(silence)} You seem nervous... You need to breathe.''}
& 3–16s 
& Before response
& 2.1\%\\
\midrule

\emph{Immediate Response}
& \texttt{Resolve} 
& When users seek information and answers directly.
& People: \textit{``...What should I do?''}
& Listener gives suggestions immediately.
& Listener: \textit{``Oh, I see. Maybe you can...''}
& 0s
& Immediate
& 29.1\%\\
\bottomrule
\end{tabular}}
\end{table*}

\subsubsection{Distribution of Pacing Strategies}
To provide quantitative guidance for CA designers on prioritizing pacing implementations, we analyzed the distribution of the observed strategies across the ten analyzed cases (total $N=289$ pacing instances). Based on the results, we computed the frequency of each strategy. As shown in \tablename{~\ref{tab:silence_strategies}}, informational and clarification strategies were predominant: \texttt{Resolve} and \texttt{Reconfirm} appeared most frequently. Conversely, high-intensity affective strategies reserved for moments of peak emotional vulnerability like \texttt{Holding} were deployed sparingly. 
This distribution was topic-dependent: for instance, the case regarding the Downward Arrow Technique (Case 2) relied heavily on clarification (15 \texttt{Reconfirm}), whereas the emotionally intense \textit{``I and Thou - touching pain and anger case''} (Case 6) prioritized deep support (5 \texttt{Holding}, 6 \texttt{Resonate}). A detailed breakdown of strategy distribution for each video is provided in \tablename{~\ref{tab:video-breakdown}} (Appendix \ref{appx:links}).

\subsubsection{Pacing Transition Patterns}
\label{subsubsec:transition}
Additionally, beyond individual response triggers, we examined the relationship between strategies to describe natural pacing shifts. We computed a transition probability matrix \cite{thomas1983sequential, cappella1980talk} that maps the likelihood of moving from one strategy to the next (ranging from 0 to 1) based on the strategy sequences observed in the video corpus (detailed in Appendix~\ref{appx:transition}). The analysis reveals two distinct patterns: Conversational Inertia and Supportive Arcs. First, we observed strong continuity (inertia) in informational and clarification contexts. As shown in Figure~\ref{fig:trans_prob_matrix}, \texttt{Resolve} and \texttt{Reconfirm} have the highest self-transition probability (0.51, 0.41), suggesting that information seeking or clarification often requires multiple turns to resolve ambiguity. Second, we identified specific ``Supportive Arcs'' for emotional regulation. Notably, the high-intensity \texttt{Holding} strategy rarely persists; instead, it frequently transitions into \texttt{Recognize} (0.40) or \texttt{Resonate} (0.40). This maps a natural therapeutic trajectory: the listener first holds space for the outburst, and then pivots to validation or emotional resonance once the intensity subsides. Conversely, \texttt{Reposition} frequently transitions to \texttt{Resolve} (0.33), suggesting that once a user's perspective is successfully reframed, they become ready to move toward solution-oriented action.

\subsection{The Design Framework for Context-Aware Pacing in Conversational Agents for Active Listening}

\subsubsection{Reflective Silence}
\emph{Reflective Silence} is mainly applicable when people share their experiences or feelings, or when they indirectly seek advice. For example, words like \textit{``I've been unsure if I'm...''} or \textit{``I just don't know if I can keep going with...''} often indicate such situations. In these cases, the listener should focus on providing emotional support and validation, helping people feel understood and giving them space to process their emerging thoughts or emotions. This type is operationalized by the \texttt{Recognize} strategy.

Prior research shows that appropriate delays can be perceived as more attentive and empathetic \cite{anzabi2023effect}, and that a ``silence followed by feedback'' rhythm gives users space to process emotions \cite{10.1145/3555164}. 
Our case analysis likewise found that listeners often keep silent briefly, through nods or short utterances like \textit{``hmm''}, before offering a full response.
Therefore, the \texttt{Recognize} strategy first offers a short emotional acknowledgment (e.g., \textit{``I see...''}) followed by a natural silence. This structure aims to convey listening and acceptance before a substantive reply.

\subsubsection{Facilitative Silence}

\emph{Facilitative Silence} applies when people express vague, contradictory, or incomplete thoughts. It is also useful when their narration fades or pauses awkwardly. Typical examples include statements like \textit{``I can't quite explain it. It's just... complicated,''} or moments of extended silence where people fail to continue.  
We operationalized this through two strategies: \texttt{Reconfirm} and \texttt{Re-engage}.

Active listening research suggests that people in such states often have not yet formed clear emotions or intentions. Gentle guidance, such as non-verbal nods or verbal prompts in the form of encouraging questions, can help them continue expressing their thoughts and feelings \cite{glenn2024so, Mc2008Teaching}. Our formative study observed two common listener tactics in this situation.
First, we observed that listeners often repeat or summarize a user’s words to help organize language and extend ideas. Based on this, we propose the \texttt{Reconfirm} strategy, which helps people clarify vague or contradictory content. It may repeat or rephrase a person's words to invite elaboration. A typical response might be: \textit{``\underline{(restatement) (silence)} Can you tell me a bit more about what you mean?''} 

Second, when a user's narration trails off, listeners often rely on non-verbal cues (e.g., eye contact and body language), or short vocalizations (e.g., \textit{``So...''} to signal encouragement. Professional listening practices describe this as ``deep listening,'' where silence before a response promotes further expression \cite{epstein2023don}. Inspired by this, we design the \texttt{Re-engage} strategy to be applied when narration fades or pauses awkwardly. It uses short verbal prompts such as \textit{``So...''} to create a guided silent space. For example: \textit{``So... Could you please explicate your actions or feelings at that time?''} This silence allows people to fill in the silence and continue, gently encouraging them to elaborate.

\subsubsection{Empathic Silence}

Whereas \emph{Reflective Silence} validates the speaker's message, \emph{Empathic Silence} addresses deeper emotional undercurrents, particularly when individuals express vulnerability or negative self-perceptions, such as \textit{``People like me just don't succeed.''} The goal here shifts from acknowledgment to active emotional co-regulation and perspective scaffolding, fostering a deeper connection. This type is operationalized by three distinct strategies: \texttt{Reposition}, \texttt{Reconsider}, and \texttt{Resonate}.

Empathic listening requires sensitivity to emotional cues and appropriate responses \cite{bodie2015sensitivity}. Studies also suggest that reflective questioning can deepen emotional involvement or help people step back from overwhelming feelings \cite{bhattacharjee2024exploring}. Our analysis aligned with this, showing two common patterns for supporting users. First, when users recalled negative experiences, listeners would confirm the emotion, briefly pause, and then redirect positively. Based on this, we introduce the \texttt{Reposition} strategy. When people share negative experiences, \texttt{Reposition} prompts them to reconsider their situation from a positive angle by combining silence with a positive reframe. For example, if a user says: \textit{``I'm always stuck,''} the listener first acknowledges the emotion, remains silent for 5–6 seconds, and then responds: \textit{``I hear you feel stuck, \underline{(silence)} but have you thought about any small changes you could try?''} Second, when users expressed self-blame, listeners instead would guide them with reflective questioning. \texttt{Reconsider} encourages them to reflect gently by combining silence with exploratory self-reflection. The listener confirms emotions, keeps silence for 2–3 seconds, and asks: \textit{``\underline{(silence)} How did that experience make you feel?''}, which avoids pressure while opening new perspectives.

Moreover, when people are immersed in long stories or detailed emotional disclosure, listeners often share a moment of silence before responding with resonance, such as expressing deep understanding or joy. Prior research suggests a correlation between the use of extended silence and increased perceptions of emotional depth and resonance \cite{atmaja2020effect}. To simulate this, we propose the \texttt{Resonate} strategy, which uses longer silences to create space for emotional connection. When people deliver long emotional narratives, \texttt{Resonate} emphasizes longer silences of 3–15 seconds, followed by responses like: \textit{``\underline{(silence)} I feel quite touched in a particular kind of way.''} This rhythm signals genuine empathy and fosters resonance.

\subsubsection{Holding Space}
\emph{Holding Space} is most applicable when people express intense negative emotions, especially intense vulnerability or distress. For instance, a user may say: \textit{``It still feels like it's all happening, like a nightmare I can't wake up from.''} Our case analysis found that continuous disclosure in such states may overwhelm people, sometimes leading to breakdowns such as crying or sobbing. If listeners continue probing, the situation may worsen. Instead, listeners often guide people to breathe deeply to reduce immediate emotional pressure. Research also shows that breathing practices, such as deep breathing, help restore calm and reduce anxiety during emotionally heavy interactions \cite{joseph2022effects}. Based on this, we propose the \texttt{Holding} strategy. The listener invites reflection and silence, guiding people to regulate emotions through breathing. For example: \textit{``\underline{(silence)} You seem nervous... You need to breathe.''} This strategy involves pauses of 3–16 seconds, providing time and space for processing emotions. The aim is to create a non-judgmental environment where people can keep silent, reflect, and experience their feelings safely.  

\subsubsection{Immediate Response}
\emph{Immediate Response} is applied when users are not seeking emotional support, but are simply seeking information or advice (e.g., \textit{``So, what should I do now?''}). While rushing to a solution may be counter-productive in supportive contexts, for task-oriented questions, an immediate response is often expected for efficiency and professionalism. This type is operationalized by the \texttt{Resolve} strategy, which delivers an answer with no deliberate silence. This approach meets users' expectations for efficiency and avoids the frustration that a redundant or delayed acknowledgment may cause in this specific, non-emotional context.

\begin{figure*}[t]
\centering
\includegraphics[width=0.7\linewidth]{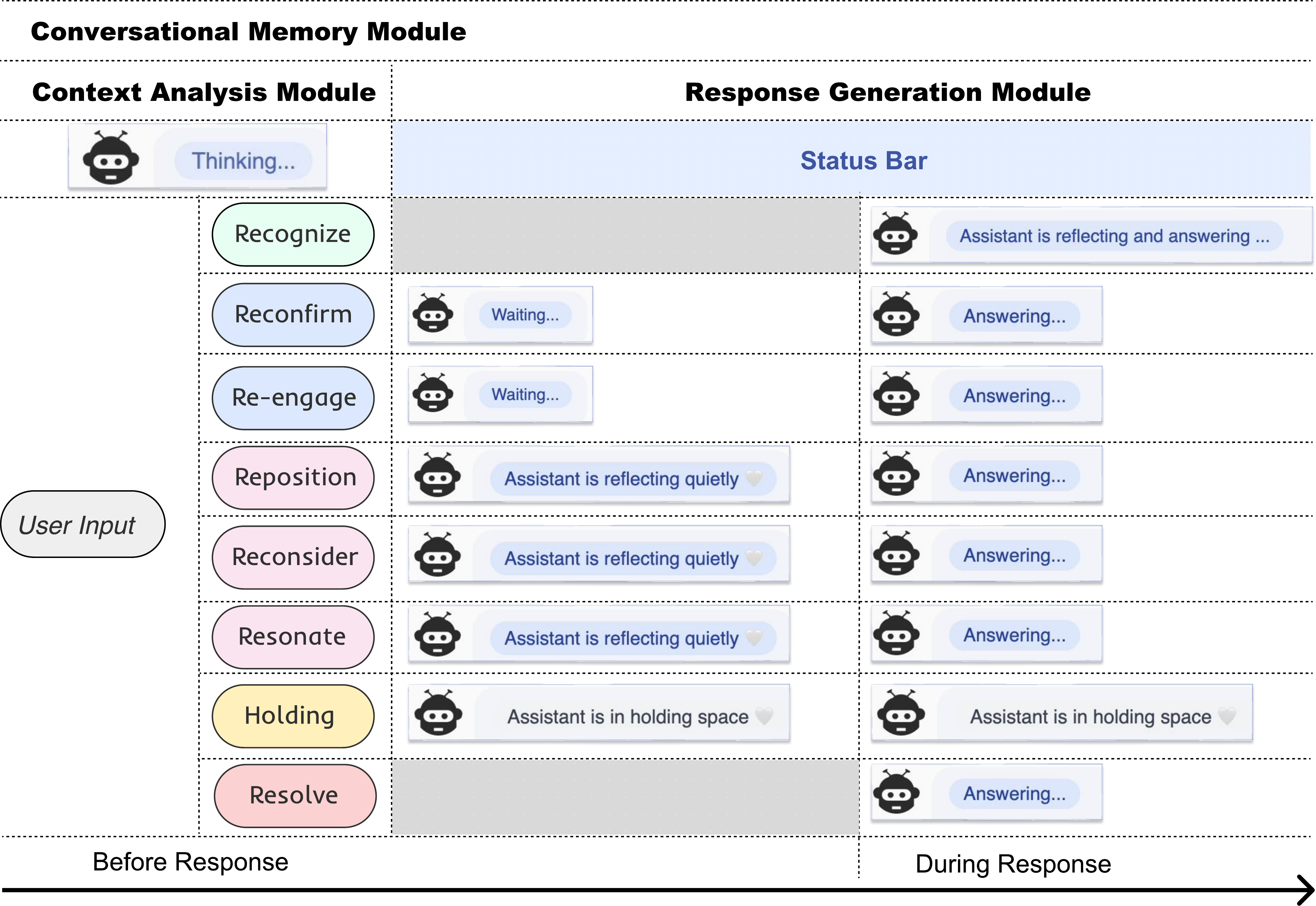}
\caption{Pipeline and visual elements of the context-aware pacing CA. It consists of three core backend modules (\emph{Context Analysis}, \emph{Response Generation}, and \emph{Conversational Memory Module}). After receiving user input, the \emph{Context Analysis Module} would first select the most appropriate pacing strategy when the status bar of the CA shows ``Thinking...''. Then its output and user input would be fed into the \emph{Response Generation Module} to generate responses and apply corresponding pacing strategies. Both modules are supported by the \emph{Conversational Memory Module} to manage the conversational contexts.}
\label{fig:ca_architecture}
\Description{Pipeline and visual elements of the context-aware pacing conversational agent (CA). This figure shows the flow of the CA system, consisting of three core backend modules: Context Analysis, Response Generation, and Conversational Memory. After receiving user input, the Context Analysis Module would first select the most appropriate pacing strategy when the status bar of the CA shows ``Thinking...''. Then its output and user input would be fed into the Response Generation Module to generate responses and apply corresponding pacing strategies. Both modules are supported by the Conversational Memory Module to manage the conversational contexts. For example, if the triggered strategy is 'Reposition', the system triggers a 'Assistant is reflecting quietly' status bar before transitioning to 'Answering...'. }
\end{figure*}

\subsection{Conversational Agent Design and Implementation}

To operationalize the findings from our formative study, we designed and implemented a context-aware pacing CA. It incorporates the five pacing types via the eight concrete strategies as listed in Table~\ref{tab:silence_strategies}. 
The CA combines a user-facing front-end with a sophisticated Python backend. The backend utilizes Flask as the web server, LangChain to orchestrate conversational logic, and the OpenAI GPT-4o API\footnote{https://platform.openai.com/docs/models/gpt-4o} for context-aware response generation. Its pipeline, as illustrated in Figure~\ref{fig:ca_architecture}, is composed of three core backend modules (\emph{Context Analysis}, \emph{Response Generation}, and \emph{Conversational Memory Modules}) that work in concert to manage conversational context and generate paced responses.
All prompts and pacing parameters are provided in Appendix~\ref{appx:prompts}.

\subsubsection{User Interface}
\begin{sloppypar}
The user interface was implemented in JavaScript as a standard chat window. 
A central design goal was to ensure the agent's context-aware pacing felt natural and non-disruptive. To avoid explicitly revealing the underlying strategies (e.g., displaying ``Empathic Silence strategy activated'') that could potentially break immersion and make the agent appear overly mechanical, we designed the visual feedback to subtly communicate the agent's processing state without exposing the mechanism. As illustrated in Fig.~\ref{fig:ca_architecture}, this was primarily achieved through a dynamic status indicator, which is a status bar that visualizes the agent's activity during silence, signaling attention rather than system lag. The specific text varied to match the agent's internal task and the function of the silence. For example, ``Assistant is reflecting quietly'' was used for Empathic Silence strategies (e.g., \texttt{Resonate}) to signal agent-side deliberation before a thoughtful response. In contrast, ``Waiting...'' was used for Facilitative Silence strategies (e.g., \texttt{Reconfirm}) as a more neutral, brief pause before offering a simple clarifying prompt. The \texttt{Holding} strategy used ``Assistant is in holding space'' both before and during the response to signal that the silence itself was the supportive action, while the \texttt{Recognize} strategy uniquely used ``Assistant is reflecting and answering...'' to show simultaneous processing and responding. These were distinct from the generic ``Thinking...'' (initial processing) and ``Answering...'' (response generation) states.\end{sloppypar}

\subsubsection{Context Analysis Module}
When a user sends a message, this module first classifies the user's intent and emotional state based on the conversational triggers identified in our formative study (see Table~\ref{tab:silence_strategies}).
This module functions as a prompt-based classifier that selects exactly one of the eight strategies (e.g., information-seeking triggers \texttt{Resolve}; high emotional valence triggers \texttt{Resonate}) shown in Table~\ref{tab:silence_strategies}. 
These eight strategies are the implementable operations that map to our five high-level conceptual types (e.g., \texttt{Resolve} is the implementation of the \emph{Immediate Response} type).
Based on the selected strategy, the module generates a control signal containing two key pieces of information: (1) the strategy label for response generation and (2) the appropriate silence duration. This silence simulates natural cognitive processing time before or during ``answering.''

\subsubsection{Response Generation Module}
The control signal from the \emph{Context Analysis Module} then conditions the \emph{Response Generation Module}. This module dynamically generates responses and adjusts behaviors based on the chosen strategy to create a natural conversational rhythm. We implement the context-aware pacing through (1) punctuation-aware micro-pauses (e.g., brief silence at commas and longer silence at ellipses) and (2) applying silence duration calculated by the \emph{Context Analysis Module} according to the corresponding timing as illustrated in Table~\ref{tab:silence_strategies}.

Specifically, for \texttt{Holding}, the stream begins with a grounding instruction (\textit{``Let's just take a deep breath here... inhale 3 seconds, exhale 3 seconds, repeat...''}) before the silence, and then continues with the generated content. Besides, when the user is inactive for 60 seconds, a separate code path triggers a brief, strategy-labeled \texttt{proactive engagement} check-in based on the chat history, such as \textit{``I'm still here if you want to continue.''}

\subsubsection{Conversational Memory Module}
To enable coherent, long-running interactions, the \emph{Conversational Memory Module} manages the dialogue history. We employ a summarization technique where a token budgeter reserves space for future replies. When the context window nears its limit, older turns are summarized into a condensed memory string, while the most recent turns are retained verbatim (e.g., keeping the last 8 turns). This ensures that the \emph{Context Analysis} and \emph{Response Generation Modules} have access to relevant historical context for making accurate, history-aware decisions.

\section{User Study}\label{sec:User Study}
To evaluate how the context-aware pacing CA impacts users' perceived quality of interaction and experience (\textbf{RQ1}) and how the context-aware pacing influences users' texting behaviors (\textbf{RQ2}), we conducted a between-subjects study under two supportive scenarios (career and relationship) compared to a static-pacing CA baseline. The study has been approved by the Institutional Review Board (IRB) and conducted ethically with university-approved IRB regulations. Participants gave consent to allow us to use their data anonymously with the understanding that the data would be deleted after the analysis.

\setlength{\textfloatsep}{4pt}
\begin{figure*}[t]
    \centering\includegraphics[width=1\textwidth]{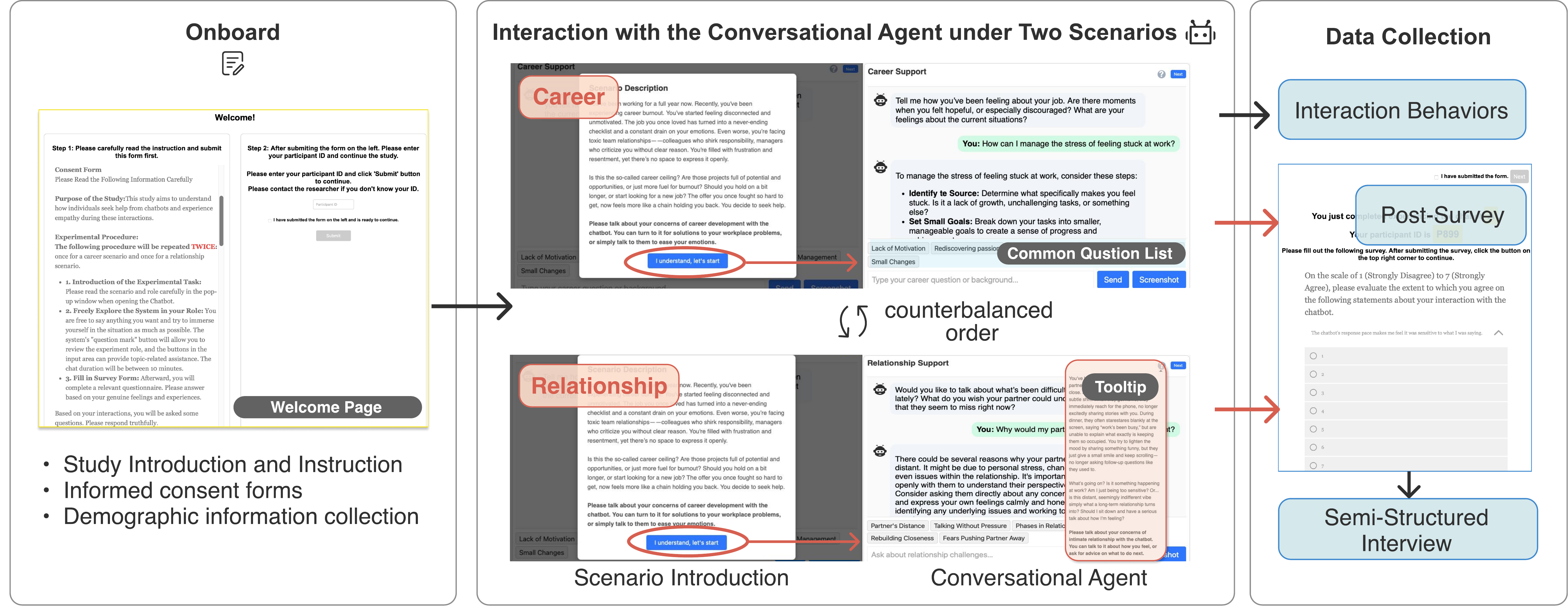}
   \caption{Overview of the user study procedure. The user first landed on the welcome page, and then interacted with the conversational agent under two scenarios, career and relationship support, in a counterbalanced sequence. After completing every scenario, the user completed a survey evaluating the corresponding experience. Finally, a semi-structured interview was conducted.}
    \label{fig:study_process} 
    \Description{Overview of the user study procedure. This figure outlines the user study procedure. The process begins with a welcome page and onboarding, which includes providing Study Introduction informed consent and collecting demographic information. The user then interacts with the conversational agent under two scenarios (career and relationship support) in a counterbalanced order. All dialogues are recorded. Upon completing all interactions, participants undertake a post-survey and a semi-structured interview.}
\end{figure*}

\subsection{Study Design}

\subsubsection{Task Scenario Description}
Participants were asked to engage in two scenarios within supportive communication. One was the career-support scenario, where participants were instructed to adopt the perspective of a young adult experiencing difficulties in career development and workplace communication. They were encouraged to seek advice from the CA on career direction uncertainty, interview anxiety, or professional skill development. Another was the relationship-support scenario, where participants sought guidance on romantic or intimate relationship difficulties, such as communication breakdowns, trust issues, or conflict resolution. These two scenarios simulated common, yet stylistically different, conversational contexts \cite{sampaio2025career, price2016young}, thereby increasing realism.

Aligned with previous HCI research \cite{10.1145/3706598.3713737, passalacqua2025safeguarding}, we utilized vignette stories to facilitate participant immersion and familiarity with the assigned topics. Prior work demonstrates that vignettes help participants adopt a protagonist's perspective and elicit real-world perceptions \cite{10.1145/3544548.3581231, lutz2021privacy, neuhaus2024mimic}, while creating a safer space to discuss private topics such as social comparisons \cite{10.1145/3706598.3713737} or help-seeking \cite{liu2022exploring}. Following this approach, all participants were invited to read two well-crafted vignette stories before starting conversations with the CA. Both stories employed second-person pronouns and offered detailed accounts of the backgrounds, problems, and feelings that participants were about to discuss. After reading the vignettes, participants were then asked to engage in the dialogues.

In the conversational stage, to ensure the diversity of topics and avoid running out of things to say, different scenarios provided participants with a ``common question list.'' This list included common topics for conversation, but participants were free to choose whether to use them to guide their discussion. 

The detailed vignette stories about scenario backgrounds and role setup, CA prompts for each scenario, and the specific common question lists can be found in Appendix~\ref{appx:scenarios}.

\subsubsection{Experimental Design}
We conducted a between-subjects experiment where 50 participants were randomly assigned to either a context-aware pacing condition or a baseline control condition. Each participant completed two supportive dialogue scenarios, with the order counterbalanced.

Given the free-form nature of user input, it was infeasible to strictly control turn-by-turn content in either condition. Instead, to ensure fair comparison, both conditions used an identical core response procedure: the same underlying model (i.e., GPT-4o), session memory mechanism, streaming transport, and visual interface. The only methodological difference was the high-level prompt strategy. While the specific strategies differed, both prompts leveraged the same high-capability base model with the shared goal of producing high-quality, supportive content.

\begin{itemize}
\item \textbf{Control group (N=25)}: Participants interacted with a static-pacing baseline CA that responded immediately, using only inherent API latency with no context-aware timing adjustments. The status bar in this condition displayed only generic states (``Thinking...'' and ``Answering...''). The baseline CA used a generic supportive prompt (Appendix \ref{appx:baseline_prompt}).
\item \textbf{Experimental group (N=25)}: Participants interacted with the designed context‑aware pacing CA that dynamically adjusted response pacing according to the user input for active listening. The context-aware CA built upon this generic supportive prompt by adding specific, dynamic pacing strategies (Appendix \ref{appx:experimental_prompt}).
\end{itemize}

As shown in Fig.~\ref{fig:study_process}, before using the CA, participants viewed a Consent Form that included the study introduction and demographic information collection. After submitting the consent, they entered the CA interface to begin the experiment.
To eliminate sequence effects, the presentation order of the two scenarios was counter‑balanced. Approximately half of the participants followed the ``Career-Relationship'' sequence, while the remainder followed the ``Relationship-Career'' sequence.
A scenario-related vignette story appeared when the interface opened and remained available via the top-right help tooltip. Additionally, the participant's typing area included a `common question list,' where they could click various questions to add them to their input box as references for topics when they were lacking inspiration. Each scenario dialogue lasted at least 10 minutes. 

\subsubsection{Participants}

\begin{table*}[t]
\centering
\small
\caption{Demographic Distribution and Chi-Square Analysis ($N=50$). The analysis confirms no significant differences between groups.}
\label{tab:demographics_chi_square}
\begin{tabular}{lccccc}
\toprule
\textbf{Demographic Variable} & \textbf{Experimental ($N=25$)} & \textbf{Control ($N=25$)} & \textbf{$\chi^2$ Statistic} & \textbf{$p$-value} & \textbf{Result} \\
\midrule
\textbf{Age} (18-24 vs. 25-34) & 14 (56\%) & 17 (68\%) & 0.3396 & 0.5601 & Not significant \\
\textbf{Gender} (Female vs. Other) & 15 (60\%) & 18 (72\%) & 0.3565 & 0.5505 & Not significant \\
\textbf{Employment} (Student vs. Other) & 19 (76\%) & 13 (52\%) & 2.1701 & 0.1407 & Not significant \\
\textbf{CA Experience} (Daily vs. Non-Daily) & 17 (68\%) & 12 (48\%) & 1.3136 & 0.2517 & Not significant \\
\bottomrule
\end{tabular}
\end{table*}

We recruited 50 participants by word of mouth and online recruitment via social media. All participants were proficient in reading and writing English, and the study was conducted in English. 
We performed the data quality check based on the following exclusion criteria: (1) technical malfunctions where the CA failed to trigger or respond; and (2) insufficient effort (e.g., participants who provided highly homogeneous or largely blank responses). All recruited participants successfully completed the study and passed data quality checks; therefore, no data was excluded.

Among them, 15 identified as males (30\%), and 33 identified as females (66\%). The remaining participants identified as non-binary or preferred not to disclose their gender. Regarding age, 31 participants (62\%) were between 18 and 24 years old, and 19 participants (38\%) were between 25 and 34 years old. As for their employment, 64\% were students, while 32\% were employed. More than half (58\%) used CAs daily. 
Participants were randomly assigned to either the context-aware experimental group or the static-pacing control group. This random assignment ensures that demographic factors and unmeasured preferences (such as personal resonance with a specific scenario) were randomly, rather than systematically, distributed. A post-hoc Chi-Square analysis confirmed that the groups were statistically equivalent: regarding age ($p=0.56$), gender ($p=0.55$), and employment status ($p=0.14$). Specifically, the experimental group ($N=25$) had 14 participants aged 18-24 (vs. 17 in control), 15 females (vs. 18 in control), and 19 students (vs. 13 in control). Prior CA experience was also broadly distributed, with only one participant in each group reporting no previous use.
As shown in \tablename{~\ref{tab:demographics_chi_square}}, given that user characteristics are statistically balanced across conditions, we inferred that the baseline propensity to resonate with the topics was equally distributed. Since the career-support and relationship-support scenarios represent common life stressors relevant to this demographic \cite{sampaio2025career, price2016young}, the likelihood to engage deeply with either topic did not differ systematically between groups. Consequently, this allowed us to attribute significant differences in the following sections to our pacing intervention rather than demographic confounds.
More details on participant demographics can be found in the Appendix~\ref{appx:participants}.

\begin{table*}[t!]
    \centering
    \footnotesize
    \caption{Questionnaire used for collecting participant feedback. All perceptual variables were rated on 7-point Likert scales.} 
    \label{tab:items}
    \begin{tabular}{llc}
    \toprule
         \textbf{Variable}&  \textbf{Item}& \textbf{Cronbach's alpha}\\\midrule
         Perceived listening&  The chatbot's response pace makes me feel it was sensitive to what I was saying.& $\alpha=0.85$\\
         &  The chatbot delivered a sense of agreement for what I was saying when appropriate.& \\
         &  The chatbot's response pace kept track of points I made.& \\
 & The chatbot assured me that it got what I said. &\\
         &  The chatbot assured me that it was receptive to my ideas. & \\
         &  The chatbot's response speed effectively conveyed a good listening attitude. & \\ 
 Human-likeness& fake/natural&$\alpha=0.855$\\
 & mechanic/humanlike&\\
 & unconscious/conscious&\\
 & artificial/lifelike&\\
 & rigid/elegant&\\ 
 Cognitive Trust& I have no reservations about acting on the chatbot's advice.& $\alpha=0.873$\\
 & I have good reason to trust the chatbot's competence. &\\
 & The chatbot handles my concerns carefully, I can confidently depend on it.&\\
 & I felt my questions were properly analyzed and addressed by the chatbot. &\\
 Affective Trust& The chatbot's response pace makes me feel it was responding to me caringly.  &$\alpha=0.803$\\
 & The chatbot displays a kind and caring attitude toward me. &\\
 & I can talk freely with the chatbot about my problems and know that it will want to listen.&\\
 Smoothness& The conversation flow was smooth. 
&$\alpha=0.879$\\
 & The response pace of the chatbot makes me feel that the conversation was natural.&\\
 &   The response pace of the chatbot makes me feel comfortable. &\\
 & The response pace of the chatbot makes me feel that the conversation was like an in-person conversation. &\\
 Interactivity& I feel that the chatbot's response pace was interactive. &\\
Manipulation Check & The chatbot pauses intentionally when responding to me.\\ 
 & The chatbot would slow down appropriately when responding to me.\\
 \bottomrule
    \end{tabular}    
\end{table*}

\begin{itemize}
    \item \textbf{Questionnaire}. After completing dialogue under each scenario, participants would fill out a questionnaire to test user experiences. We also conducted manipulation checks. All measures were rated on a 7-point Likert scale (1 = Strongly Disagree, 7 = Strongly Agree). The variables include perceived listening \cite{10.1145/3555164}, human-likeness \cite{bartneck2009measurement}, cognitive and affective trust \cite{johnson2005cognitive}, smoothness \cite{10.1145/2212776.2223750}, interactivity \cite{10.1145/2317956.2318069}. Details of all measurement items are provided in \tablename{~\ref{tab:items}}.
    \item \textbf{Interview}. After completing both scenarios, to better understand how participants interpret CA's response pacing during the study, we conducted semi-structured interviews with all participants \cite{adeoye2021research}. We also analyzed participants' responses to the three open-ended questions in the post-survey. We specifically asked how and why participants perceived CA's response pace differently (See Appendix~\ref{appx:interview} for the interview outline).
    \item \textbf{Interaction Behaviors}. Participants' interactions with the CA during the experiment, including the conversation content, timestamp of each conversation, and pacing strategies applied (if applicable), were recorded.
\end{itemize}

\subsection{Data Analysis Approach}

\subsubsection{Questionnaire}

We collected participants' perception of (a) perceived listening, (b) human-likeness, (c) trust (cognitive and affective), (d) smoothness, and (e) interactivity. We calculated Cronbach's $\alpha$ to assess the internal consistency of the measurement items (\tablename{~\ref{tab:items}}). Mann-Whitney U tests were conducted to compare the responses from the Experimental group and the Control group.

\subsubsection{Interview Thematic Analysis}
We employed an iterative, consensus-based thematic analysis \cite{hammer2014confusing} on data from post-experiment interviews and open-ended questionnaire responses. Following transcription, four researchers independently performed open coding to identify themes regarding context-aware pacing. Through collaborative discussions and preliminary exercises, we developed an initial codebook, regularly refining codes to ensure reliability until theoretical saturation was achieved. The initial set of codes covered perceptions of context-aware pacing (``response speed,'' ``active listening content,'' etc.), conversational behaviors in different scenarios (e.g., ``seeking emotional support.''); emotional experiences during conversations (e.g., ``need satisfaction''), etc. Finally, we iteratively refined these into core themes by examining contextual meanings and eliminating irrelevant codes.

\subsubsection{Interaction Behaviors}

To evaluate how the CA's context-aware pacing influenced participants' texting behaviors, we quantified their depth of self-disclosure and engagement levels based on their conversation contents. The depth of self-disclosure was measured using the total emotion words, first-person pronoun usage, and user turn length:
    \begin{itemize}
        \item Total Emotion Words. We leveraged the NRC Word-Emotion Association Lexicon, which contains over 14,000 English words and associates them with eight basic emotions based on Plutchik's wheel, as well as positive and negative sentiment \cite{perrie2013using}. We used this library to sum the total count of word occurrences in the user input that are associated with an emotion or sentiment. This metric reflects affective self-disclosure, which is the act of sharing personal feelings and emotional states \cite{Studer2025The,Liu2022The, reuel-etal-2022-measuring}. We posit that a higher total emotion word count signifies a greater volume of emotional disclosure.
        \item Total First-Person Pronoun Usage. We defined lists of first-person singular pronouns (i.e., ``I'', ``me'', ``my'', ``myself'', and ``mine'') and plural pronouns (i.e., ``we'', ``us'', ``our'', ``ours'', and ``ourselves'') and counted their usage in the user's input. This count serves as a linguistic marker of self-focused attention and self-disclosure, indicating that a speaker is referencing their own experiences, thoughts, or feelings \cite{Brockmeyer2015Me, reuel-etal-2022-measuring}.
        \item User Input Length (in Words). Measuring the word count of user messages is a common method for estimating the depth of self-disclosure, based on the assumption that longer, more detailed responses reflect greater engagement and willingness to share personal information \cite{Barak2007Degree, Lee2022Online, Pan2018What}.
    \end{itemize}
The user engagement level is measured using the number of conversation turns and pacing strategy distribution:
\begin{itemize}
    \item Total Conversation Turns \cite{shum2018eliza}. We counted the total number of conversational exchanges, with one turn defined as a user input followed by a CA response. We used this as our primary metric for engagement, based on the assumption that more engaged users will continue the dialogue for a greater number of turns \cite{Jiang2023Weakly, Zhou2018The, Venkatesh2018On}.
    \item Pacing Strategy Distribution. For the experimental group, we further analyzed the frequency distribution of the different pacing strategies applied by the CA. This distribution serves as a proxy for the nature and diversity of user engagement. A more diverse distribution indicates a deeper, multifaceted engagement.
\end{itemize}
Mann-Whitney U tests were conducted to compare the depth of self-disclosure and engagement levels of the Experimental group and the Control group.

\section{Results}\label{sec:Result}
In the following sections, we present the results of our quantitative and qualitative analyses. For all Mann-Whitney U tests, we report the U-statistic ($U$) and p-value ($p$), and the rank-biserial correlation coefficient ($r$) which serves as a measure of effect size.

\subsection{Manipulation Check}

To better examine whether our manipulations of context-aware pacing were successfully designed, we asked two questions about their perceptions of pacing and the appropriateness of pacing (see the survey questionnaire in \tablename{~\ref{tab:items}}). The two-sided Mann-Whitney U Test showed that our manipulations were successful across the two scenarios (career: $U=56$, $p<0.001$, $r=0.71$; relation: $U=64$, $p<0.001$, $r=0.7$). Participants in the experimental group perceived a significant pacing change (career: $M=5.68$, $SD=1.25$; relation: $M=5.4$, $SD=1.00$) compared to those in the control group (career: $M=2.92$, $SD=1.5$; relation: $M=3.24$, $SD=1.39$). The experimental group also rated these changes (career: $M=5.12$, $SD=1.05$; relation: $M=5.20$, $SD=1.08$) as significantly more appropriate (career: $U=78.5$, $p<0.001$, $r=0.65$; relation: $U=78.5$, $p<0.001$, $r=0.68$) than the control group (career: $M=2.96$, $SD=1.46$; relation: $M=3.04$, $SD=1.46$).

    \begin{table*}[t!]
        \centering 
        \small
        \renewcommand{\arraystretch}{1}
        \captionof{table}{The accuracy and distribution of pausing strategies. Specifically, the Career/Relationship-Accuracy represents the number (and percentage) of strategies that were classified correctly by the CA.}
        \label{tab:strategy}
        \begin{tabular}{lccccc}
            \toprule
            \bfseries Pausing Strategy & \bfseries All & \bfseries Career & Career-Accuracy & \bfseries Relationship & Relationship-Accuracy\\
            \midrule
            \texttt{Holding} & 8 & 5 (2.3\%) & 5 (100\%) & 3 (1.4\%) & 3 (100\%)\\
           \texttt{Resolve} & 94 & 49 (22.7\%) & 46 (93.9\%) & 45 (21.4\%) & 39 (86.7\%)\\
           \texttt{Recognize} & 70 & 33 (15.3\%) &30 (90.9\%) & 37 (17.6\%) & 28 (75.7\%)\\
           \texttt{Reconfirm} & 103 & 47 (21.8\%) & 41 (87.2\%) & 56 (26.7\%) & 44 (78.6\%)\\
           \texttt{Reconsider} & 35 & 16 (7.4\%) & 13 (81.3\%) & 19 (9.0\%) & 14 (73.7\%)\\
           \texttt{Reposition} & 60 & 38 (17.6\%)& 36 (94.7\%) & 22 (10.5\%)& 18 (81.8\%)\\
           \texttt{Resonate} & 15  & 5 (2.3\%)& 4 (80.0\%) & 10 (4.8\%) & 7 (70.0\%)\\
           \texttt{Re-engage} & 41 & 23 (10.6\%) & 21 (91.3\%) & 18 (8.6\%) & 18 (100\%)\\
           \midrule
           \bfseries Overall & 426 & 216 (50.7\%) & 196 (90.7\%) & 210 (49.3\%)  & 171 (81.4\%)\\
            \bottomrule
        \end{tabular}
        \end{table*}

\subsection{Semantic Content Analysis for Confound Check}
\label{subsec:semantic}
As discussed in the previous section, the inherent variability in the turn-by-turn content generated by CAs made it infeasible to strictly control for content. Therefore, to determine whether there are any underlying differences in the semantic content of the agents' responses between the experimental and control conditions that could confound our findings, we performed a semantic content analysis on the full corpus of agent responses from both groups.

We employed the all-MiniLM-L6-v2 sentence-transformer model\footnote{\url{https://huggingface.co/sentence-transformers/all-MiniLM-L6-v2}} to generate dense vector embeddings for all agent responses, enabling quantitative comparison of semantic content. For each scenario, we computed pairwise cosine similarities within each condition (intra-group) and between conditions (inter-group) \cite{reimers2019sentence}. 
To assess potential content shifts, we adopted a distributional analysis approach analogous to cluster validity estimation \cite{rousseeuw1987silhouettes}.
We posited that if pacing strategies fundamentally altered the content of the advice, the semantic similarity between conditions (inter-group) would be significantly lower than the natural semantic variation observed within a single condition (intra-group).
We computed pairwise cosine similarities for each scenario and tested for divergence using the Mann-Whitney U test and Jensen-Shannon divergence for distribution overlap \cite{lin2002divergence}.

Results demonstrate strong semantic similarity across both scenarios, with no significant differences found between conditions.
For the career scenario, there is no significant difference between the mean within-experimental similarity (0.203) and between-group similarity (0.199), as indicated by a Mann-Whitney U test ($p=0.440$) and a negligible effect size ($r=0.004$). The relationship domain showed a similar pattern, with a mean within-experimental similarity of 0.250 and between-group similarity of 0.255.
Statistical analysis showed no significant divergence ($p=0.494$, $r=0.004$). In both cases, low Jensen-Shannon divergence values (career: 0.033; relationship: 0.085) confirmed that the semantic content distributions were highly overlapping.

These findings suggest that, despite variations, the experimental and control groups exhibit similar semantic content. This check for confounds demonstrates that the perceptual differences reported in the following sections can be more reliably attributed to our primary manipulation of pacing strategies.
        
\subsection{Context Classification Performance}
\label{sec:performance_evaluation} 
Since our architecture implicitly synthesizes user intent and emotional cues to drive strategy selection, we evaluated the module's reliability based on the accuracy of the final deployed strategy. This metric serves as a holistic proxy for the system's understanding of the conversational context. We conducted a post-hoc evaluation using the interaction logs collected during the user study. Two researchers independently annotated these turns based on the coding scheme defined in \tablename{~\ref{tab:silence_strategies}}, categorizing the ideal pacing strategy (e.g., \texttt{Resolve}, \texttt{Recognize}, \texttt{Holding}) for each context. Inter-rater reliability was calculated using Cohen's Kappa (career: $\kappa = 0.9$; relation: $\kappa = 0.93$), indicating strong agreement between the two coders. 
Any discrepancies between the two coders were resolved through discussion to establish a final consensus-based ground truth.
We then compared the ground-truth annotations against the actual strategies selected by the Context Analysis Module during the study. The system achieved a high overall accuracy of 86.2\%. The accuracy of each strategy can be found in \tablename{~\ref{tab:strategy}}.
Performance was robust across both scenarios, though higher in the Career scenario (90.7\%) compared to the Relationship scenario (81.4\%). Specifically, the module showed strong precision in identifying task-oriented intents requiring Immediate Response (\texttt{Resolve} accuracy: 90.4\%) and conversational stalls requiring \texttt{Re-engage} (95.1\%). Strategies for critical emotional states also performed well, with \texttt{Holding} strategy achieving 100\% accuracy and \texttt{Reposition} achieving 90.0\%. However, performance was relatively lower for strategies requiring deeper semantic interpretation of subtle emotional cues, such as \texttt{Resonate} (73.3\%) and \texttt{Reconsider} (77.1\%). A qualitative review of these error cases suggests they were primarily benign misclassifications between semantically similar supportive strategies (e.g., substituting \texttt{Reconsider} for \texttt{Resonate}), rather than critical failures. The results confirm that the pacing strategies were deployed appropriately according to the experimental design.

 \begin{figure*}
          \centering
          \subfloat[Listening]{
            \centering
            \includegraphics[width=.15\textwidth]{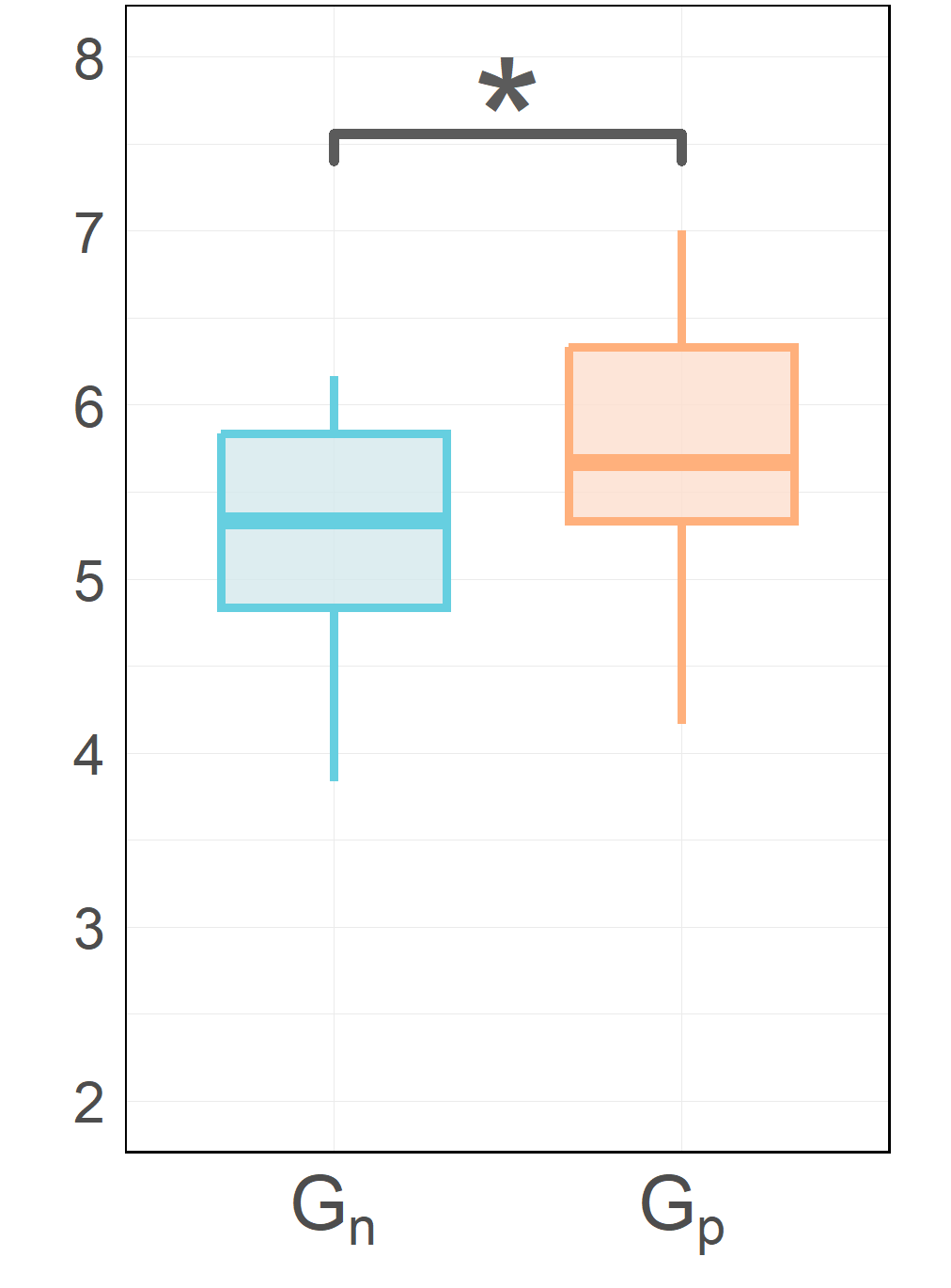}
            \label{fig:listening}
          }
          \subfloat[Affective Trust]{    
                  \centering
                  \includegraphics[width=.15\textwidth]{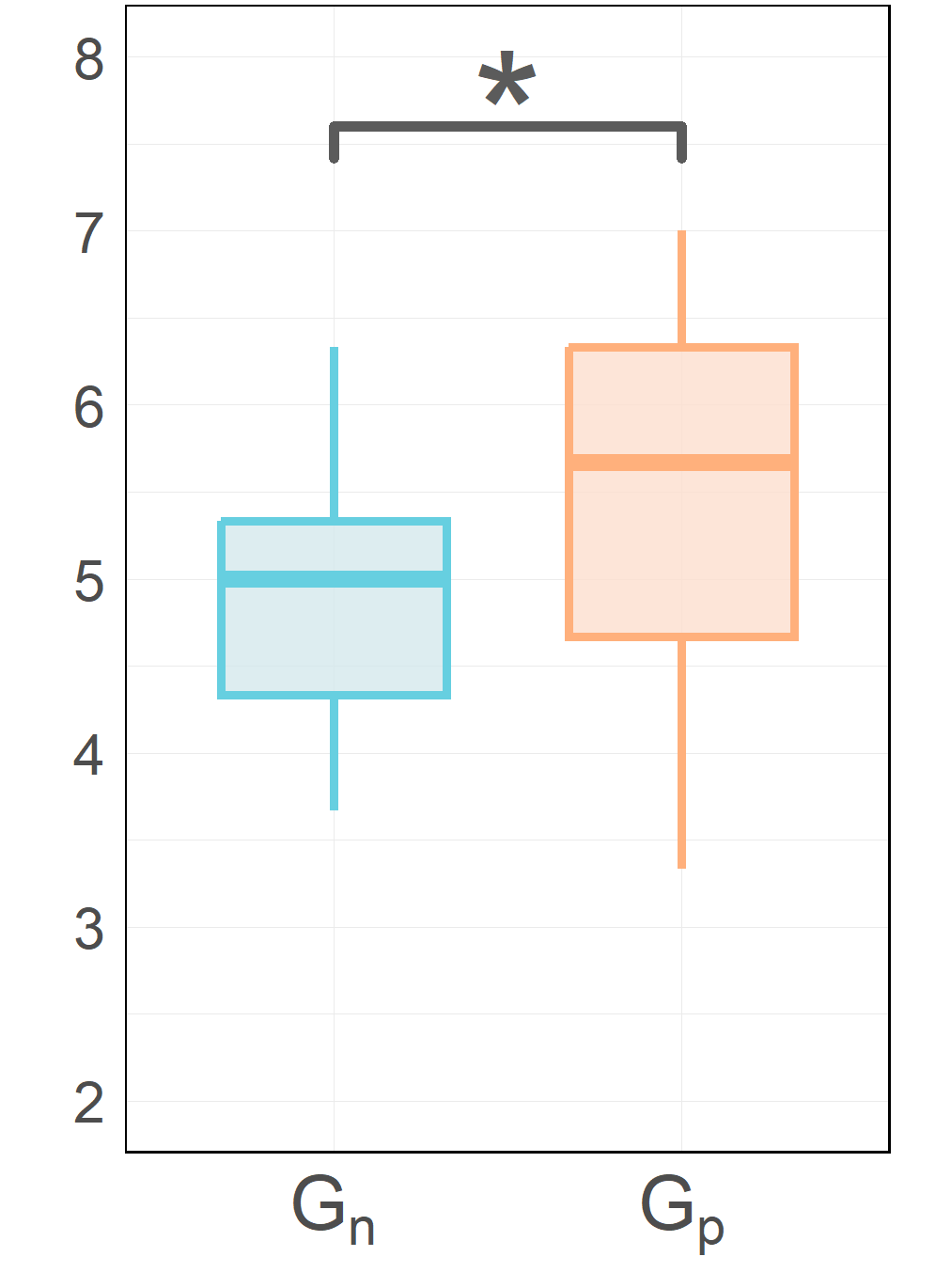}
              \label{fig:Affective Trust}
          }
          \subfloat[Cognitive Trust]{
                  \centering
                  \includegraphics[width=.15\textwidth]{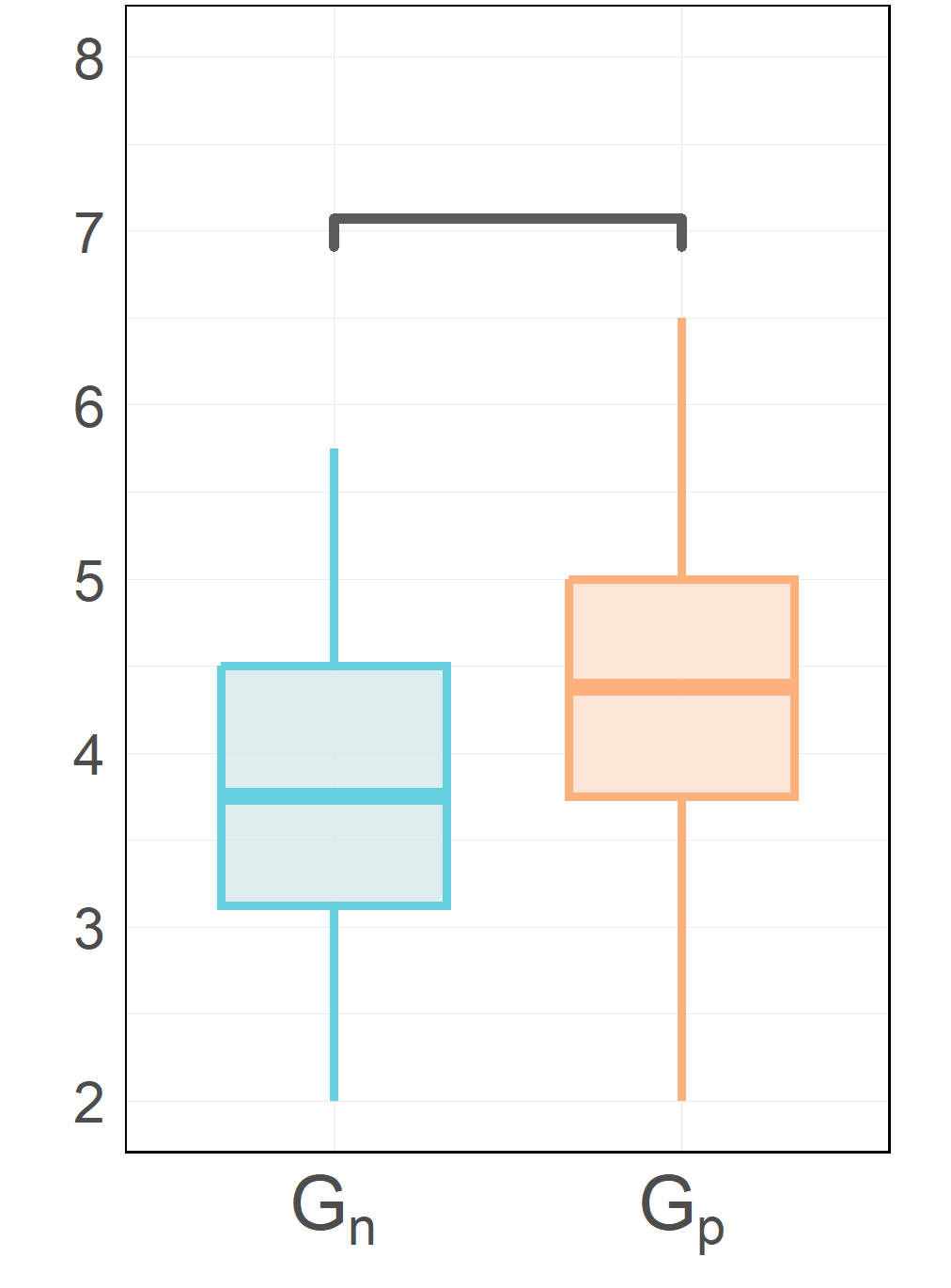}
              \label{fig:cognitive trust}
          }
          \subfloat[Human-likeness]{    
                  \centering
                  \includegraphics[width=.15\textwidth]{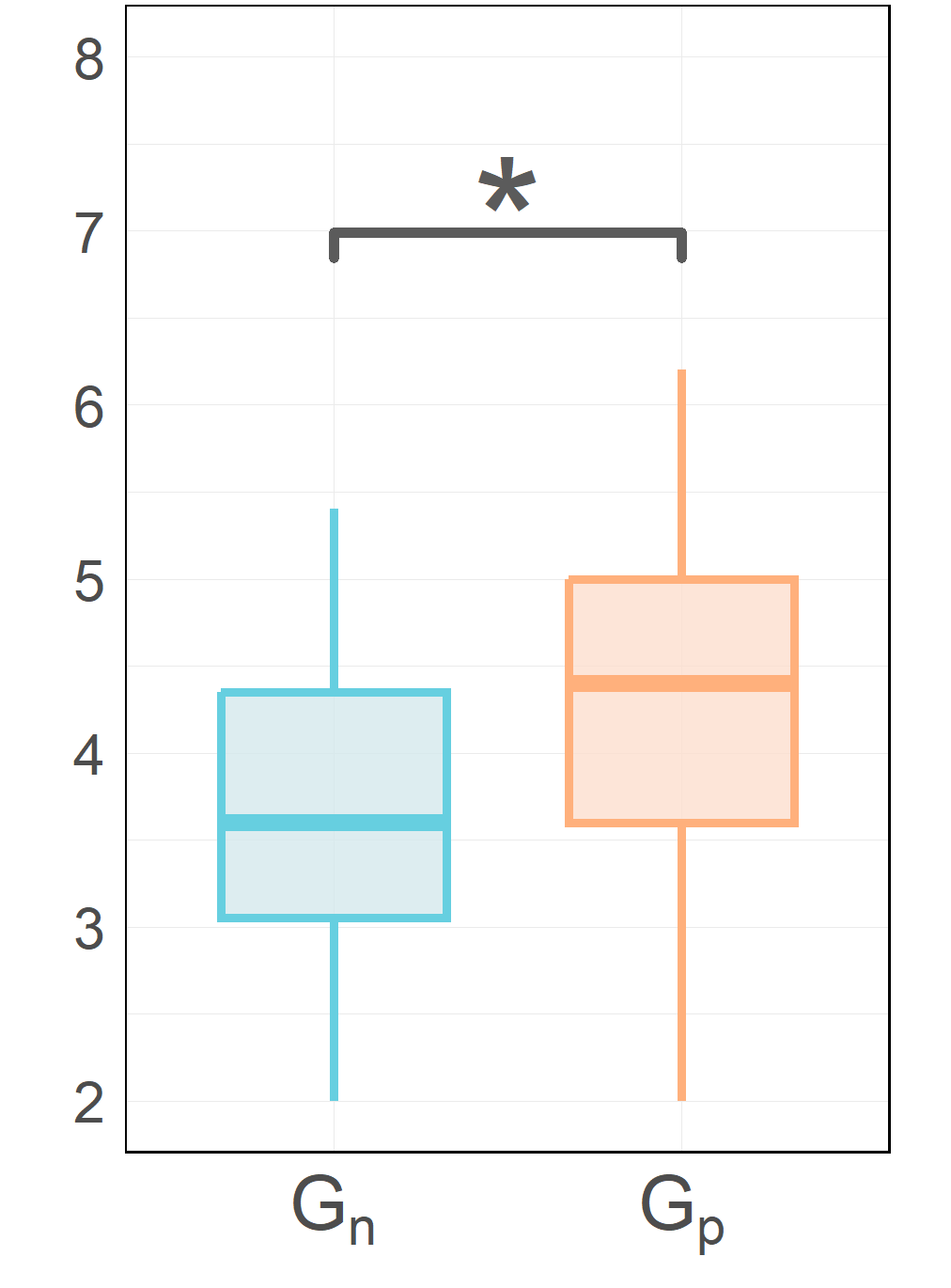}
              \label{fig:human-likeness}
          }
            \subfloat[Smoothness]{    
                  \centering
                  \includegraphics[width=.15\textwidth]{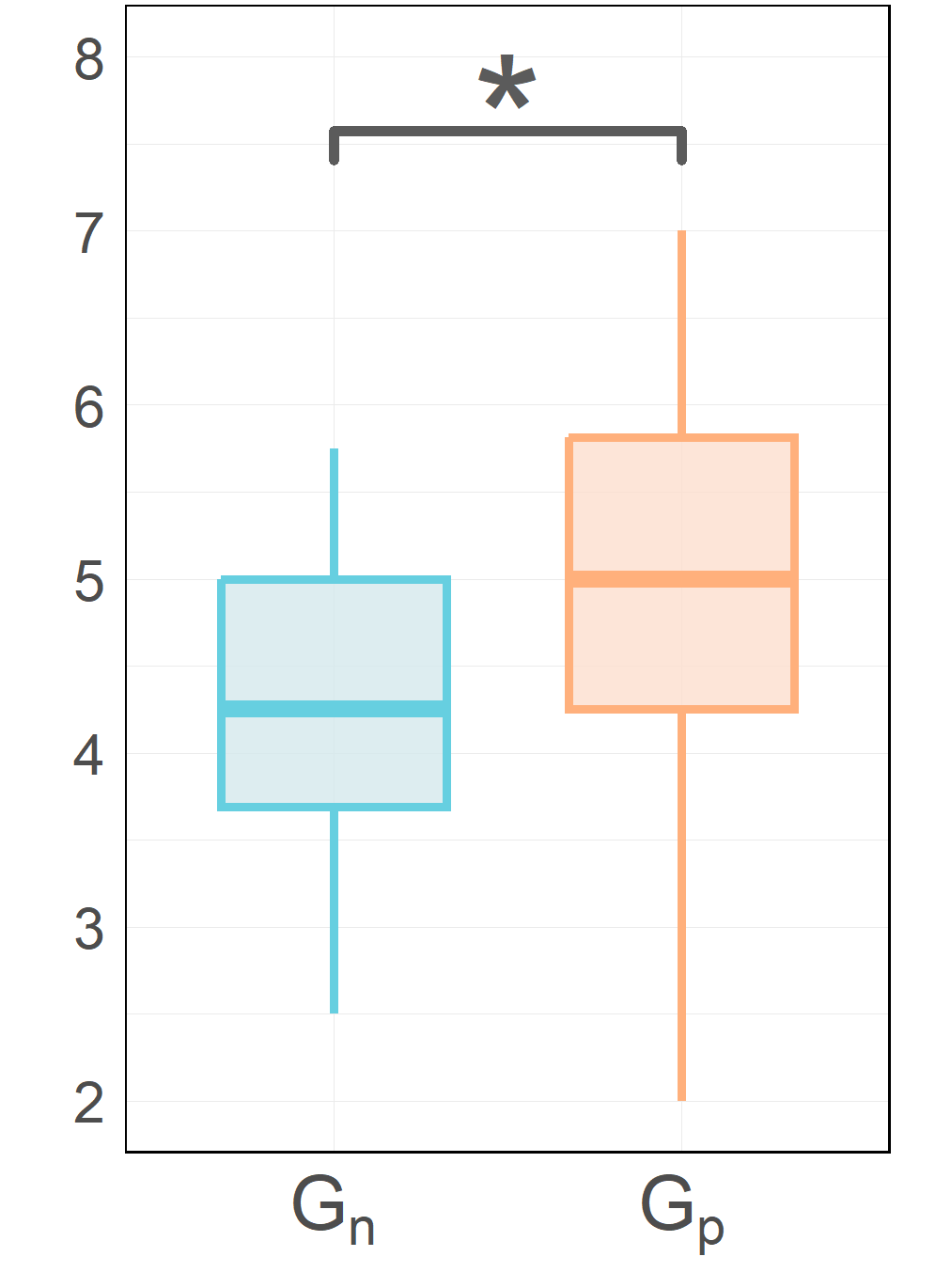}
              \label{fig:smoothness}
          }
          \subfloat[Interactivity]{
                  \centering
                  \includegraphics[width=.15\textwidth]{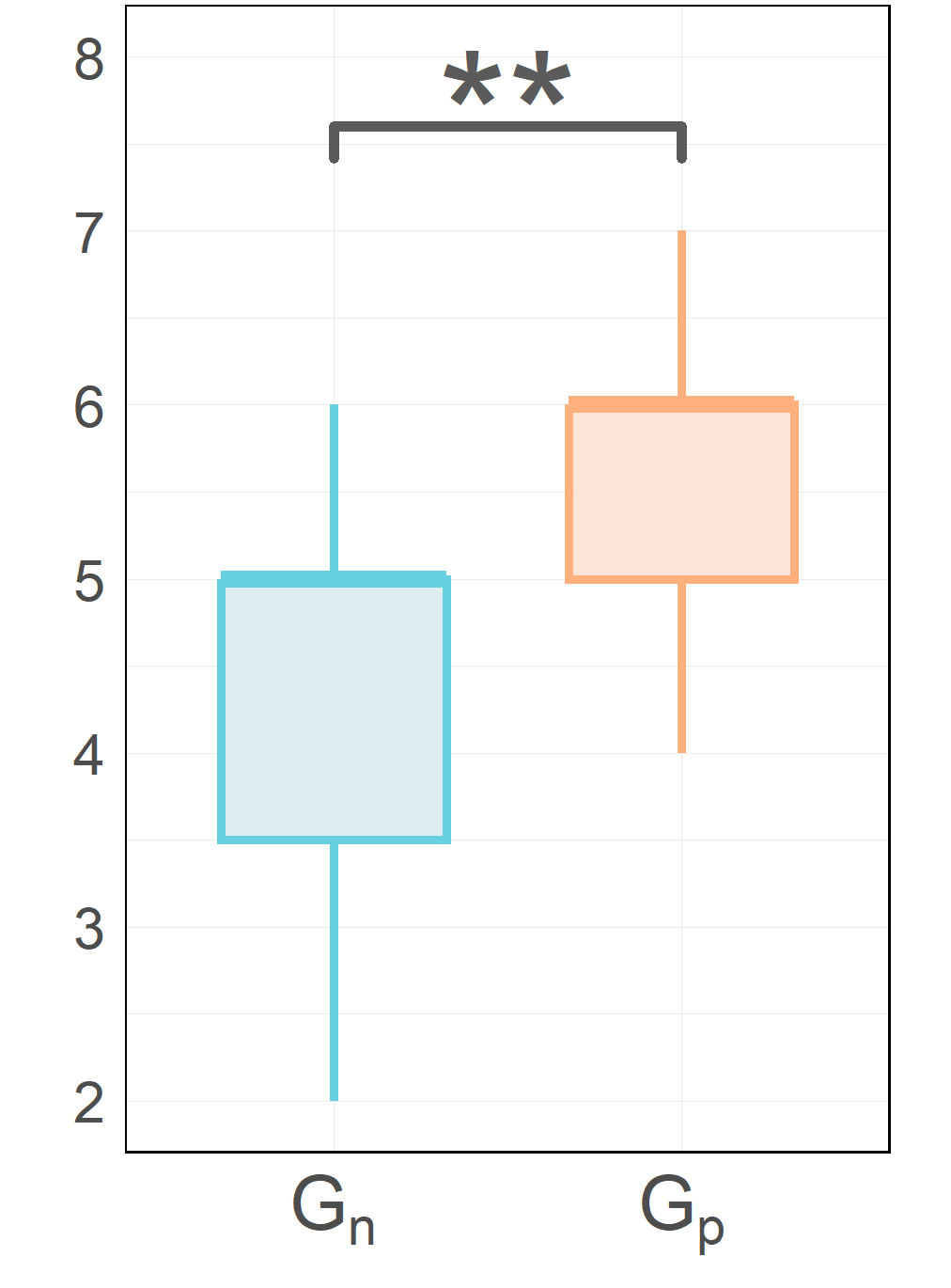}
              \label{fig:Interactivity}
          } 
        \caption{Statistical analysis on questionnaire results in career-support scenario, with $G_n$ referring to the control group, and the context-aware pacing group $G_p$. These figures show the significant difference between the control and experimental group in terms of perceived listening quality, affective trust, human-likeness, smoothness, and interactivity.}
          \label{fig:survey-career}
          \Description{Career-support scenario questionnaire results for control and experimental groups.This figure presents the statistical analysis of questionnaire results from the career-support scenario, comparing the control group and the experimental group with context-aware pacing. The data indicate that the experimental group demonstrated significant improvements over the control group across multiple metrics: perceived listening quality (M = 5.74 vs. M = 5.05, SD = 0.89; U = 188, p = 0.016, r = 0.34), affective trust (M = 5.45 vs. M = 4.73; U = 197, p = 0.024, r = 0.32), human-likeness (M = 4.26 vs. M = 3.42; U = 182, p = 0.011, r = 0.36), smoothness (M = 4.80 vs. M = 4.14; U = 205, p = 0.024, r = 0.30), and interactivity (M = 5.56 vs. M = 4.08; U = 141, p = 0.001, r = 0.48).}
\end{figure*}
\begin{figure*}
          \centering 
        \subfloat[Listening]{    
                  \centering
                  \includegraphics[width=.15\textwidth]{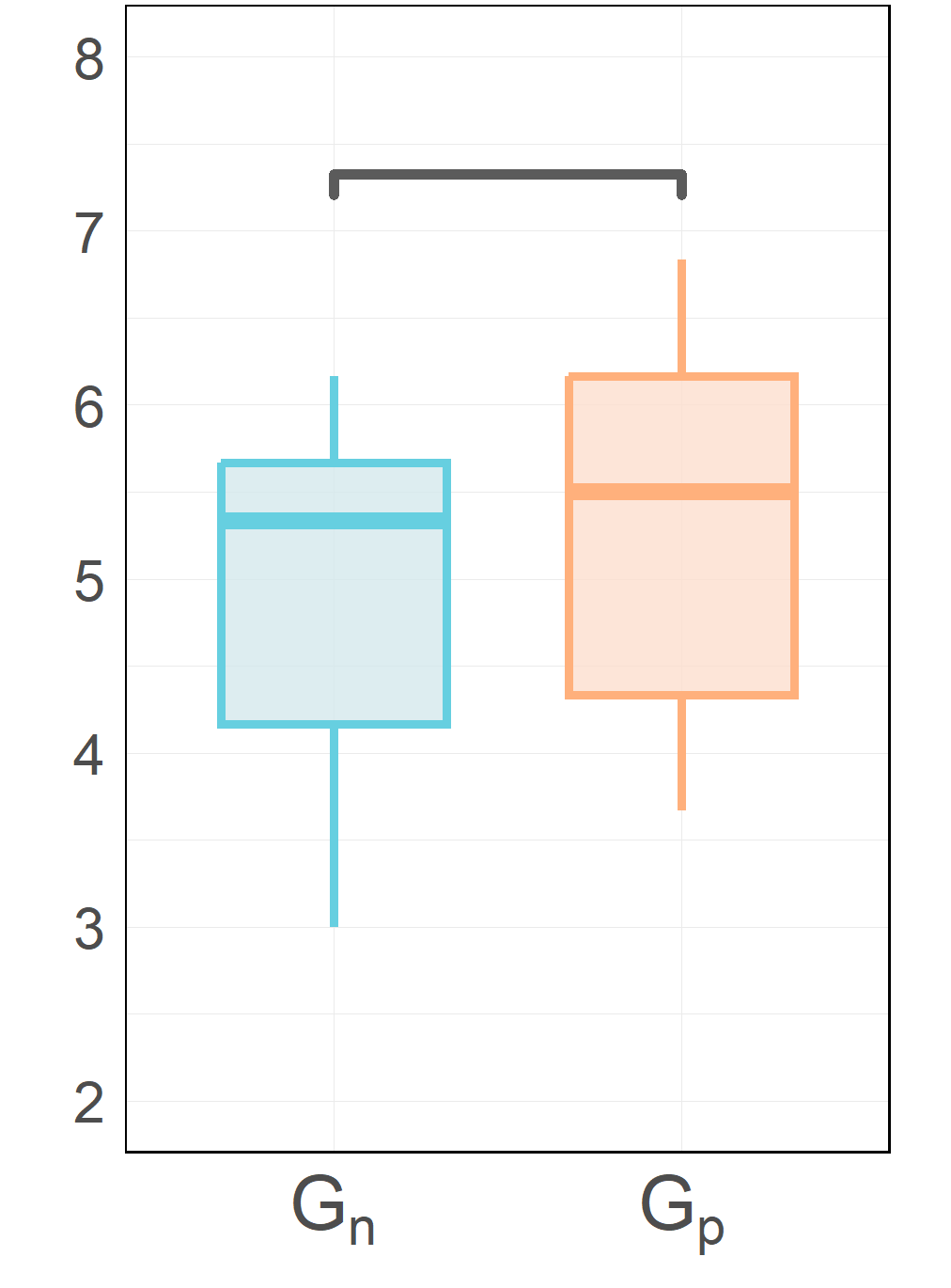}
              \label{fig:listening_B}
          }
          \subfloat[Affective Trust]{    
                  \centering
                  \includegraphics[width=.15\textwidth]{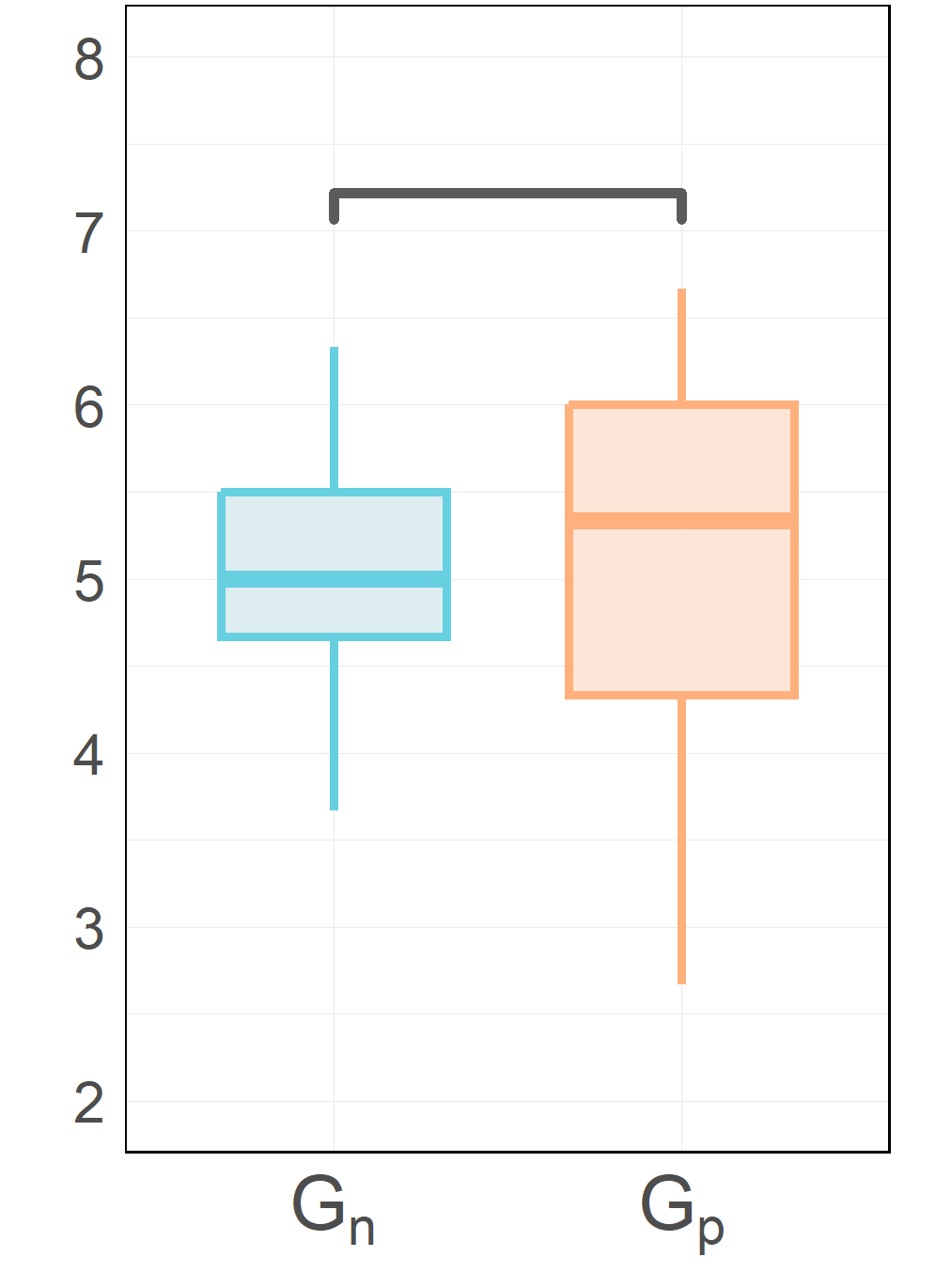}
              \label{fig:affective trust_B}
          }
          \subfloat[Cognitive Trust]{    
                  \centering
                  \includegraphics[width=.15\textwidth]{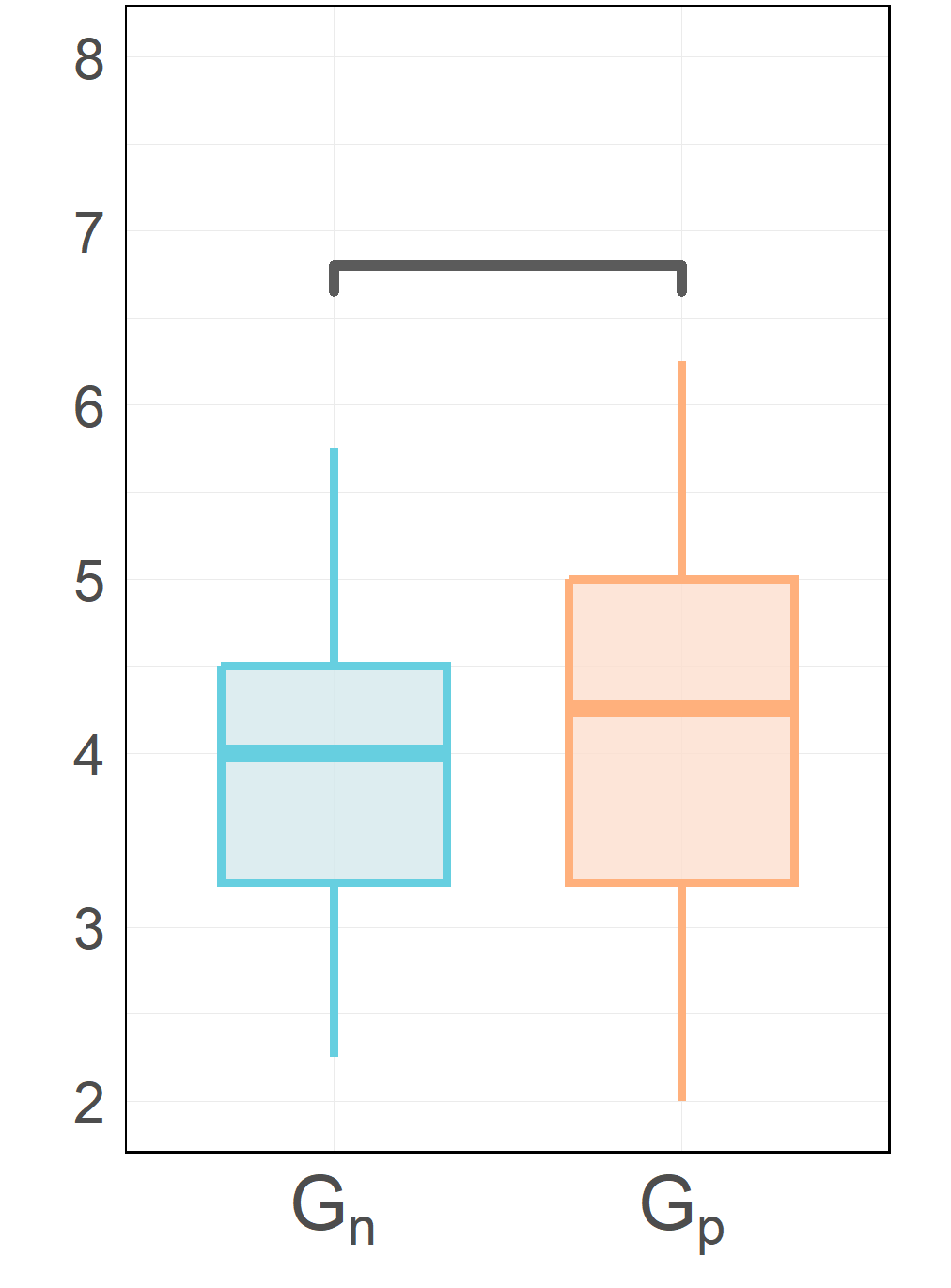}
              \label{fig:cognitive trust_B}
          }
          \subfloat[Human-likeness]{    
                  \centering
                  \includegraphics[width=.15\textwidth]{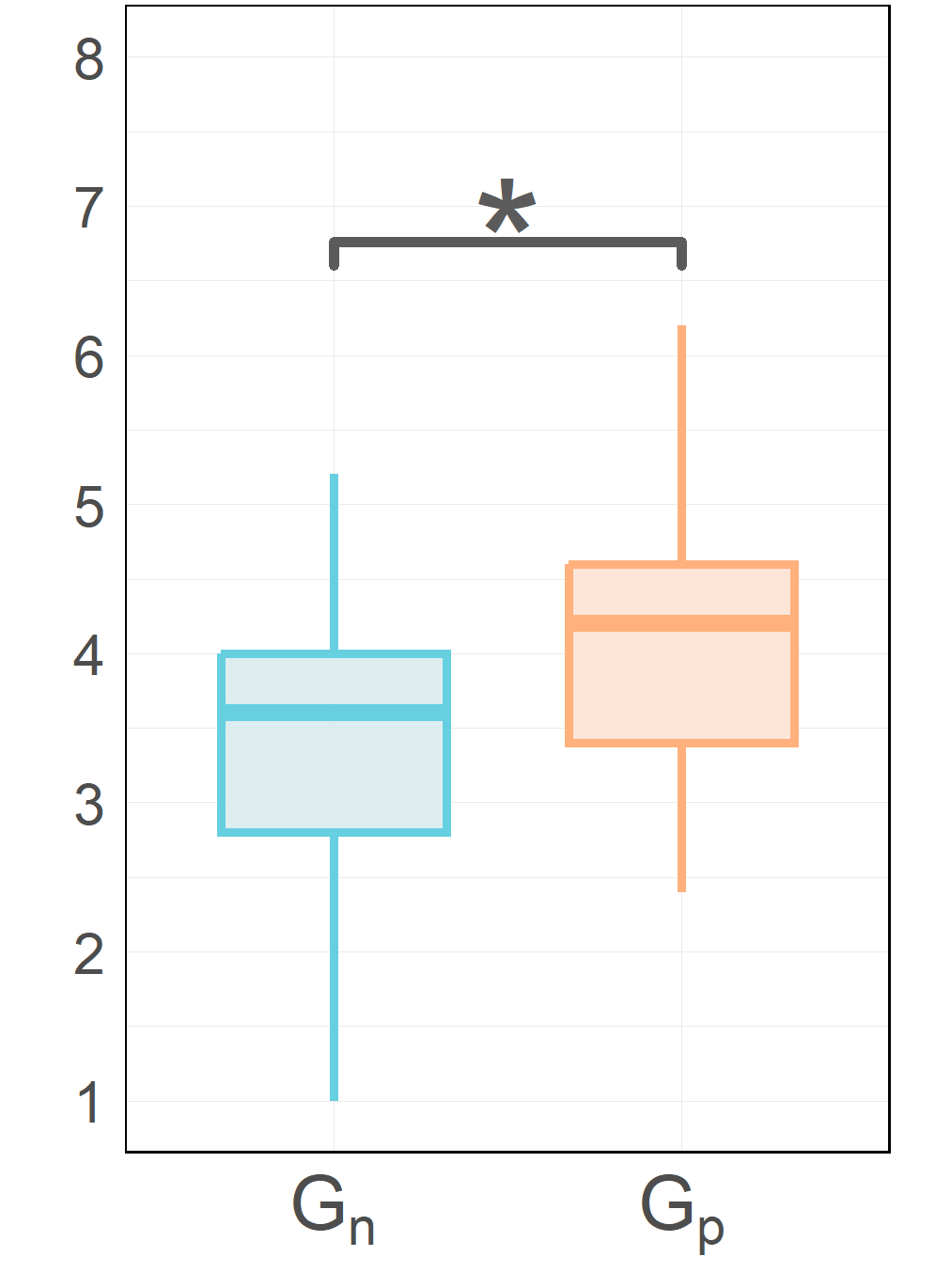}
              \label{fig:human-likeness_B}
          }
          \subfloat[Smoothness]{    
                  \centering
                  \includegraphics[width=.15\textwidth]{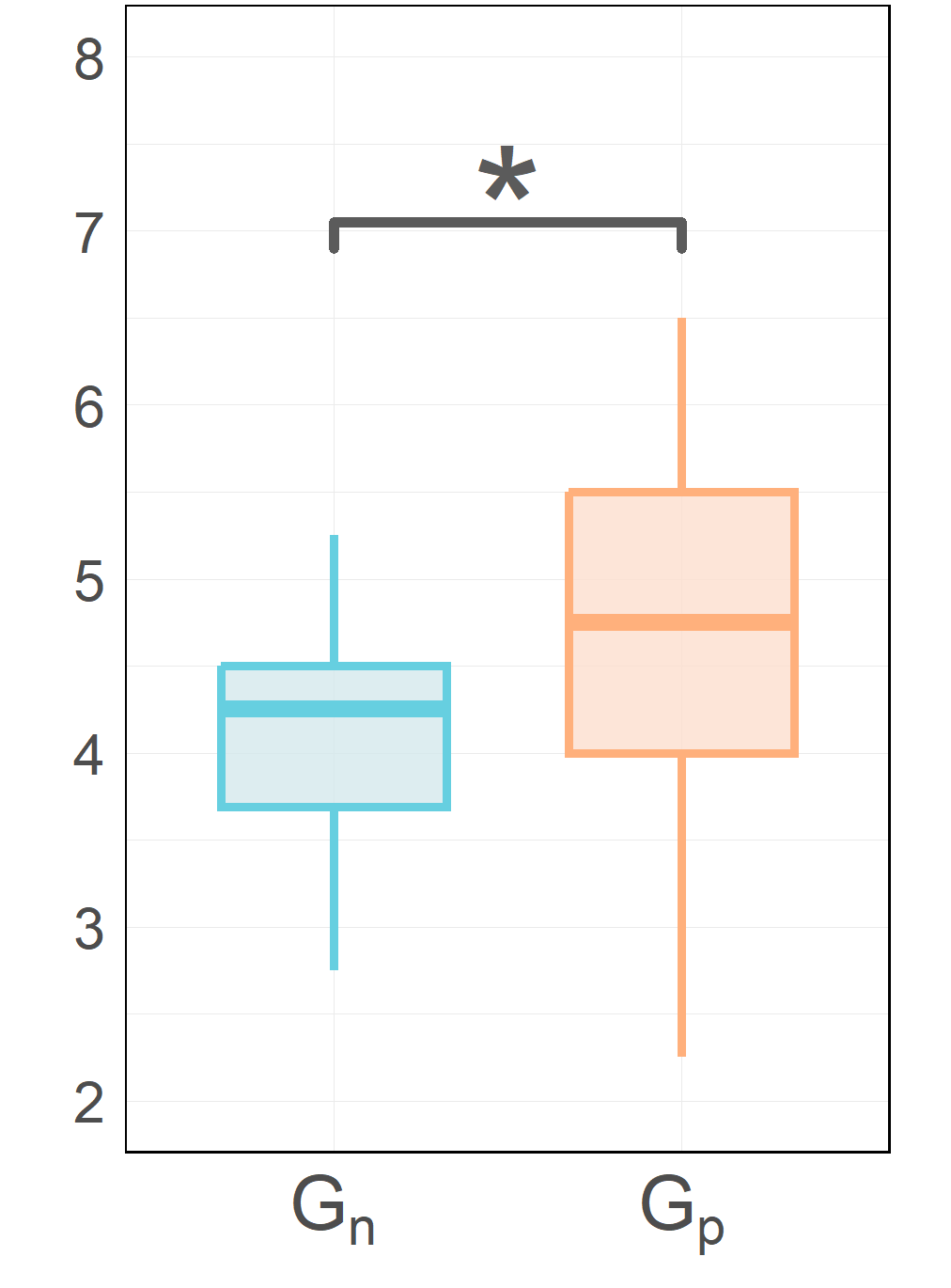}
              \label{fig:smoothness_B}
          }
          \subfloat[Interactivity]{    
                  \centering
                  \includegraphics[width=.15\textwidth]{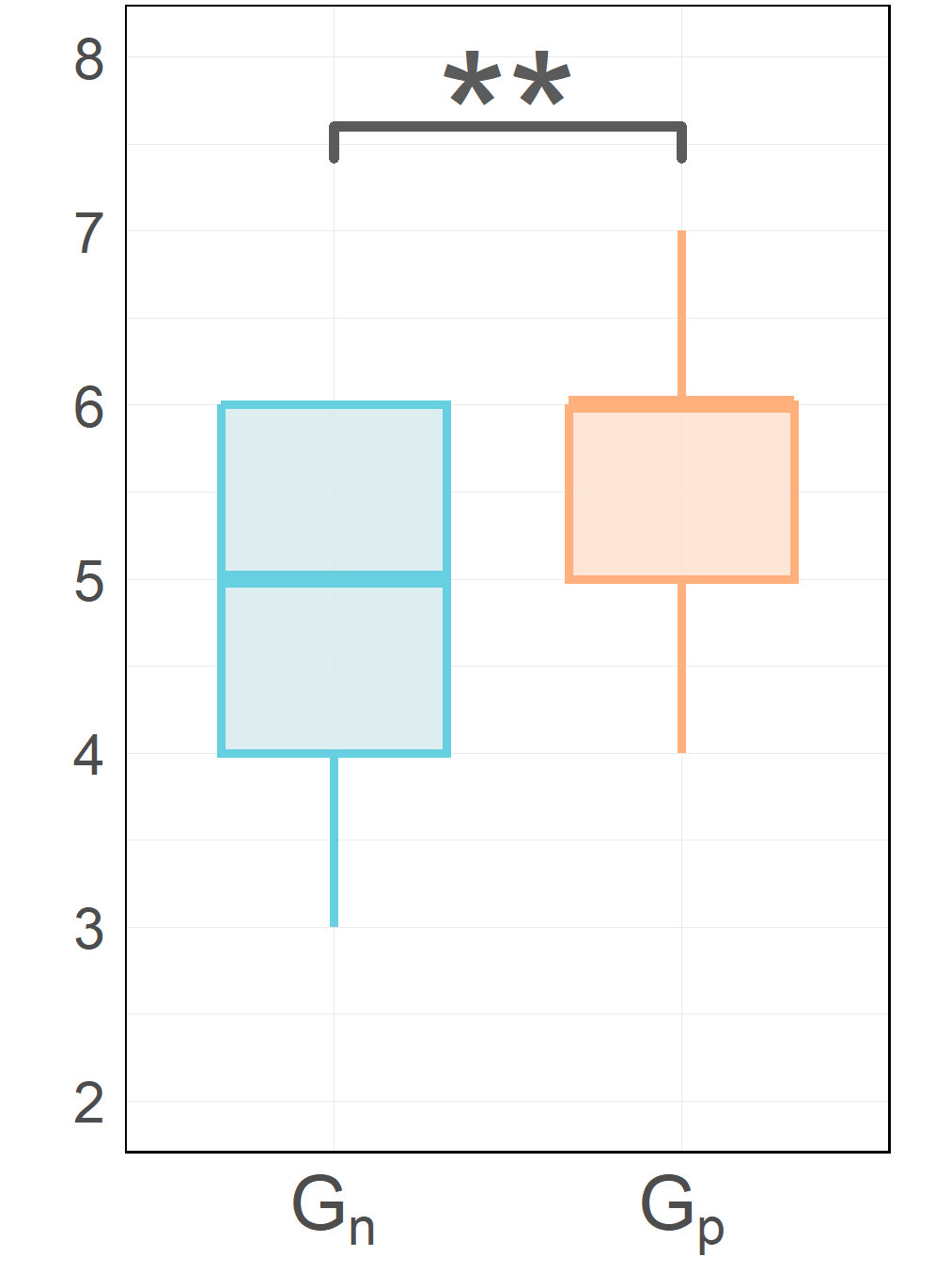}
              \label{fig:Interactivity_B}
          }
         
          \caption{Statistical analysis on questionnaire results in relation-support scenario, with $G_n$ referring to the control group, and the context-aware pacing group $G_p$. These figures show the significant difference between the control and experimental group in terms of perceived human-likeness, smoothness, and interactivity.}
          \label{fig:survey-relation}
          \Description{Relationship-support scenario questionnaire results for control and experimental groups. This figure presents the statistical analysis of questionnaire results from the relationship-support scenario for both the control group and the experimental group with context-aware pacing. Results show that the experimental group significantly outperformed the control group in human-likeness (M = 4.16 vs. M = 3.36; U = 206, p = 0.039, r = 0.30), smoothness (M = 4.73 vs. M = 3.99; U = 180, p = 0.010, r = 0.36), and interactivity (M = 5.64 vs. M = 4.56; U = 162, p = 0.002, r = 0.43).}
      \end{figure*}

\subsection{The Impact of Context-Aware Pacing on Perceived Interaction Quality and User Experiences (\textbf{RQ1})}

To evaluate how context-aware pacing influenced user perception, we analyzed survey data using two-sided Mann-Whitney U tests and situated these findings within qualitative interview feedback. We compared outcomes for participants using the context-aware agent (experimental group) with those using a static-pacing agent (control group) across two scenarios: career support and relationship support.

\subsubsection{Pacing Increased Affective Trust and Listening Quality in Career-Support Scenario by Simulating Deliberation}
\label{subsubsec:aff_lis}
A key finding was that context-aware pacing significantly improved relational metrics, though this effect was most pronounced in the career-support scenario. Participants rated the context-aware agent significantly higher on perceived listening quality in the career scenario ($M=5.74$, $SD=0.89$; $U=188$, $p=0.0158$, $r=0.343$) compared to the control agent ($M=5.05$, $SD=1.06$), as shown in Figure~\ref{fig:survey-career}(a). The context-aware pacing also significantly improved affective trust in the career-support scenario (Figure~\ref{fig:survey-career}(b); $M=5.45$ vs. $M=4.73$; $U=197$, $p=0.024$, $r=0.32$).
These effects were not statistically significant in the relationship scenario (Figure~\ref{fig:survey-relation}), nor was cognitive trust significantly affected in either scenario.

\begin{figure*}
    \centering
    \includegraphics[width=1\linewidth]{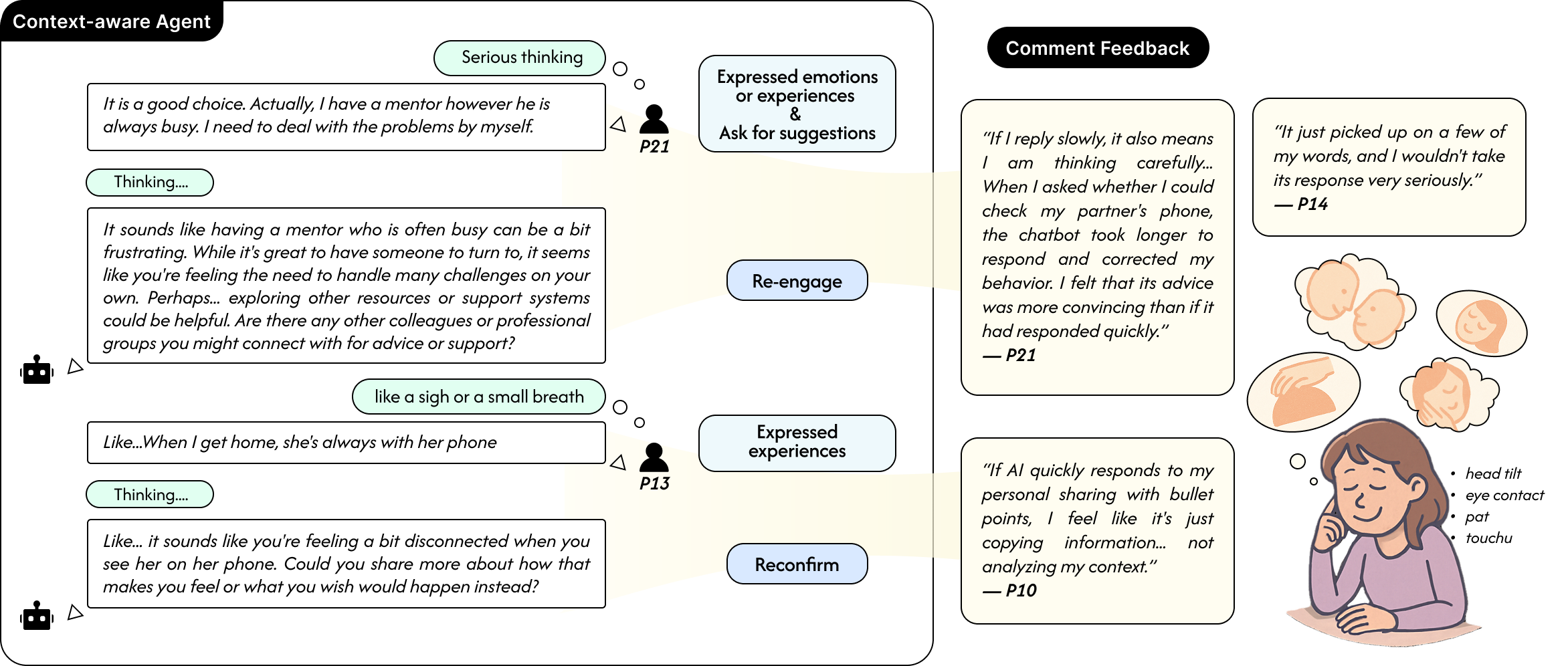}
    \caption{Pacing enhanced both persuasion and human-likeness by simulating a more deliberate interaction, boosting trust and listening quality, while creating a stronger sense of social presence. This made the interaction feel more natural and human-like.}
    \label{fig:good}
    \Description{Pacing Enhances Interaction Quality and Social Presence in Human-Agent Communication. The figure illustrates how deliberate response pacing in Conversation Agent interactions increases persuasion and human-likeness, using participant quotes as examples. Slower response (e.g., P21: "If I reply slowly, it also means I am thinking carefully") was perceived as more thoughtful, boosting trust and listening quality. Conversely, fast, generic answers (e.g., P10: "quickly responds... with bullet points, I feel like it's just copying information") reduced perceived authenticity. This pacing strategy created a stronger sense of social presence, making interactions feel more natural and addressing user concerns about superficial responses.}
\end{figure*}

Qualitative feedback explains how context-aware pacing built affective trust: participants interpreted silence as evidence of cognitive effort and care, as shown in Figure~\ref{fig:good}. This perception of ``thinking'' enhanced the credibility of the agent's advice and made it feel more supportive. Participants frequently used terms like \textit{``serious thinking''} (P21), \textit{``deliberation''} (P14), and \textit{``special attention''} (P6) to describe this. P21 drew a direct analogy to human behavior: \textit{``If I reply slowly, it also means I am thinking carefully... When I asked whether I could check my partner's phone, the chatbot took longer to respond and corrected my behavior. I felt that its advice was more convincing than if it had responded quickly.''} Conversely, rapid responses from the control agent were sometimes perceived as superficial, undermining trust. P14 noted, \textit{``It just picked up on a few of my words, and I wouldn't take its response very seriously.''}

This feedback, which links perceived deliberation directly to trust, also helps explain the specificity of our quantitative findings. We interpret this scenario-specific effect as being driven by the nature of the task. The career-support scenario is a hybrid task that blends the need for practical, actionable advice with emotional validation. In this context, the agent's ``deliberation'' signaled that it was ``seriously thinking'' about the user's problem, which enhanced the affective trust and listening quality. This cue was less relevant in the more purely affective relationship scenario.
This also explains the lack of effect on cognitive trust, which relates to perceived competence and accuracy. The pacing didn't make the agent seem smarter (it was the same GPT-4o model), but it did make it seem more caring and less superficial, thus boosting affective trust specifically in the context where thoughtful advice was most valued.

\subsubsection{Pacing Enhanced Human-Likeness by Creating a Sense of Social Presence}
Context-aware pacing could enhance participants' human-likeness perception in both career-support scenario ($U=182$, $p = 0.011$, $r=0.36$) and relationship-support ($U=206$, $p = 0.039$, $r=0.30$), as shown in Figure~\ref{fig:survey-career}(d) and Figure~\ref{fig:survey-relation}(d). Participants explained that the temporal variations made the agent feel less like a machine and more like an attentive entity, as shown in Figure~\ref{fig:good}. The silence successfully simulated conversational backchannels and even physical presence. For example, participants described silence as being \textit{``like a sigh or a small breath''} (P12) or \textit{``like a faint breath in conversation... makes me feel the other person is there''} (P10).
Remarkably, participants extended this feeling of presence to imagined physical embodiment. P8 described silence as resembling a friend's comforting \textit{``pat''} or \textit{``touch''}, while P10 imagined a \textit{``head tilt''} or \textit{``eye contact''}. In contrast, the static control agent's speed was often described as distancing and mechanical: \textit{``If AI quickly responds to my personal sharing with bullet points, I feel like it's just copying information... not analyzing my context.''} (P10)

\subsubsection{Pacing Improved Overall Interaction Quality}
The intervention significantly improved general metrics of interaction quality across both scenarios. As illustrated in Figure~\ref{fig:survey-career}(e) and Figure~\ref{fig:survey-relation}(e), smoothness ratings were higher for the context-aware agent in both the career ($U=205$, $p = 0.024$, $r=0.3$) and relationship scenarios ($U=180$, $p = 0.010$, $r=0.36$). Similarly, perceived interactivity saw a strong improvement (Figure~\ref{fig:survey-career}(f) and Figure~\ref{fig:survey-relation}(f)) in both career ($U=141$, $p = 0.001$, $r=0.48$) and relationship scenarios ($U=162$, $p = 0.002$, $r=0.43$). This suggests that dynamic pacing made the conversation flow feel more natural overall.

\subsubsection{The Double-Edged Sword of Pacing: Violating Machine Efficiency Heuristics}
\label{subsubsec:doubleedged}
\begin{figure*}
    \centering
    \includegraphics[width=1\linewidth]{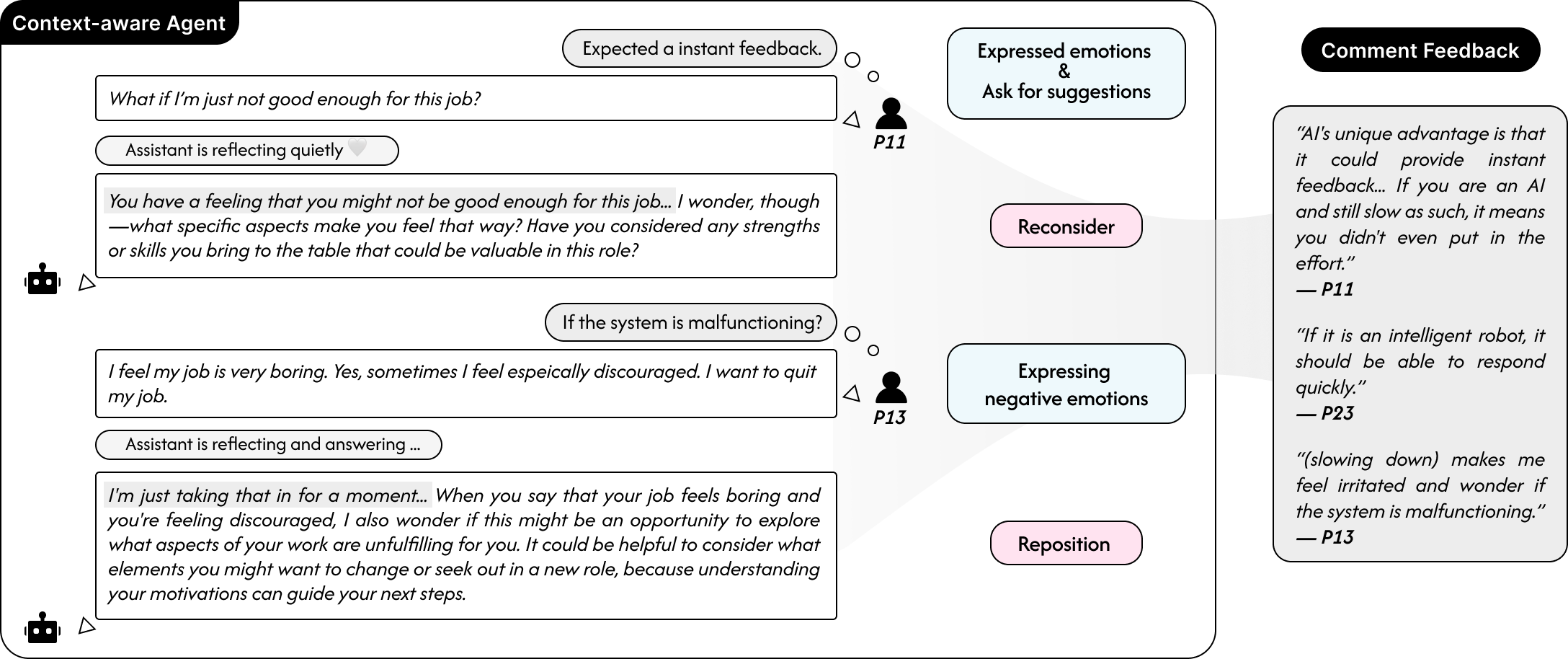}
    \caption{Some participants expressed that they would prefer a quick response over slow-paced emotional support. Otherwise, it feels more like a malfunction of the AI than an empathetic strategy.}
    \label{fig:instant}
    \Description{Participant Expectations for Rapid AI Response. The figure illustrates how active listening strategies in intelligent robot interactions are perceived as malfunctions rather than intentional design, using direct participant quotes as examples. Slow pacing (e.g., P13: "makes me feel irritated and wonder if the system is malfunctioning") was interpreted as a technical failure. Conversely, immediate feedback was explicitly expected from an intelligent system (e.g., P23: "it should be able to respond quickly"), with delays leading to frustration and the perception that the AI was not putting in effort (P11).}
\end{figure*}

Despite its widespread benefits, pacing was not universally positive. For a notable minority of participants, slowness violated the machine heuristic, namely the expectation that AI should be faster and more efficient than humans \cite{10.1145/3290605.3300768} (N11, P18). This led to frustration, particularly when participants felt the agent's response quality did not justify the wait. Participants expected high utility from AI, and slow responses could be perceived as poor performance or system malfunction.
\textit{``AI's unique advantage is that it could provide instant feedback... If you are an AI and still slow as such, it means you didn't even put in the effort.'' }(P11)
Some participants interpreted the slower pacing as poor model capability, leading to a sense of disengagement (P8) and negative expectancy violation, as shown in Figure~\ref{fig:instant}. \textit{``If it is an intelligent robot, it should be able to respond quickly.'' (P23)} Their default expectation is that the robot will respond quickly and give lengthy answers. \textit{``(slowing down) makes me feel irritated and wonder if the system is malfunctioning.''} (P13) Even though our status bar indicates that the slower pacing means the chatbot ``thinking'' or ``reflecting,'' some participants still automatically assumed it was merely an excuse for loading issues (P23). These negative interpretations suggest that while slower pacing can convey listening and connection, it may also evoke users' concerns of inefficiency, asynchronicity, and uncertainty.

In summary, the results demonstrate that context-aware pacing significantly enhances users' interaction quality and experience with CAs, with consistent improvements across both supportive scenarios in perceived human-likeness, smoothness, and interactivity. Affective trust and perceived listening quality were also significantly enhanced, although this effect was primarily concentrated in the goal-oriented career-support context. This positive perception was driven by interpreting pacing as thoughtful deliberation. However, this effect was in tension with some users' expectations of machine efficiency, creating a design tension between perceived empathy and perceived performance.

\begin{figure*}[!t] 
        \centering 
        \includegraphics[width=.8\linewidth]{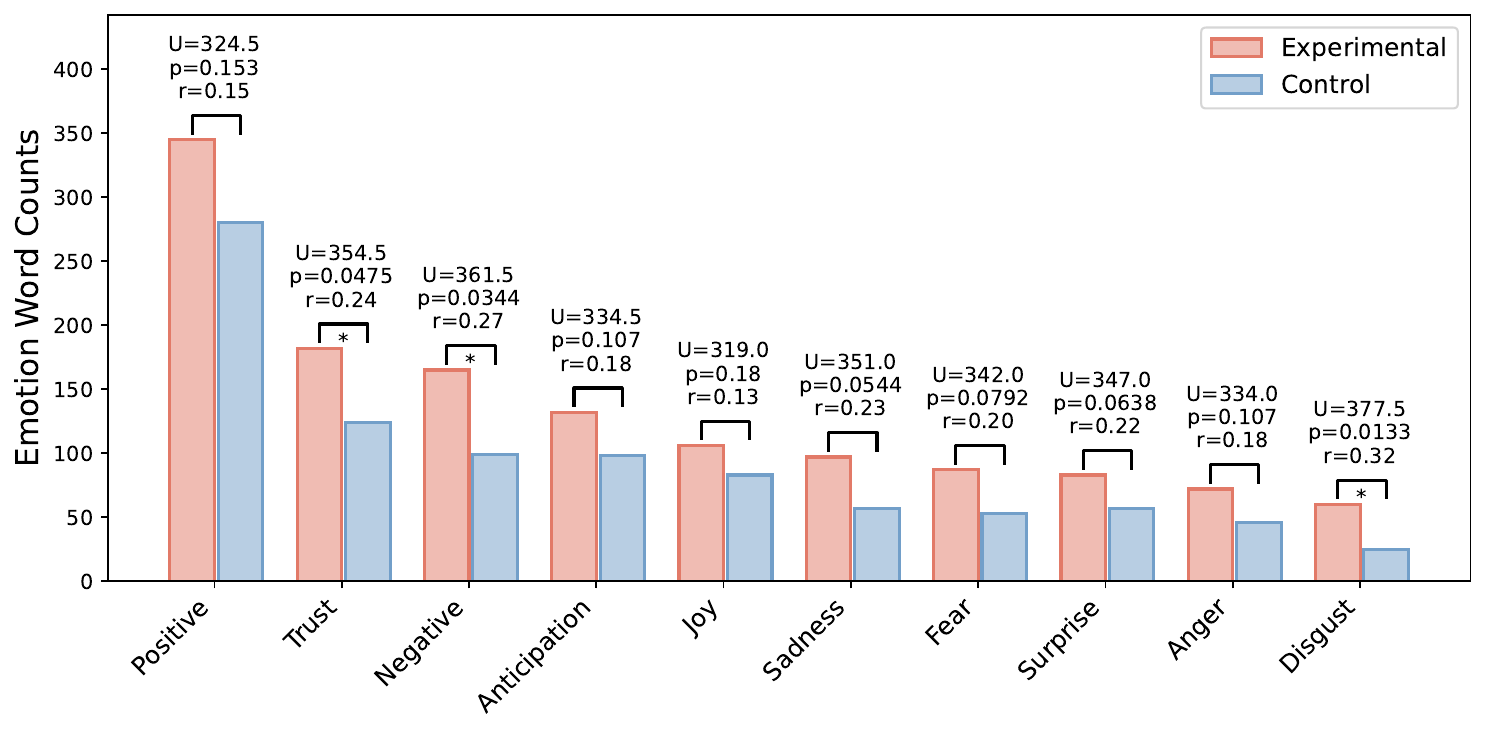} 
        \vspace{-0.8cm}
        \caption{The emotion word count distribution across different groups. $U$ is the test statistic. $p$ is the $p$-value. $r$ is the effect size.}
        \label{fig:emotion}
        \Description{Emotion word count distribution across experimental and control groups. The bar graph compares the frequency of emotion words between the Experimental and Control groups across ten categories: Positive, Trust, Negative, Anticipation, Joy, Sadness, Fear, Surprise, Anger, and Disgust. The data reveals a consistent trend where the Experimental group exhibits higher word counts than the Control group in every category. Statistical annotations indicate significant differences (p < 0.05) for Trust, Negative, and Disgust emotions, with marginal significance (p < 0.1) observed for Sadness, Fear, and Surprise. The "Positive" category shows the highest overall volume of words for both groups, while "Disgust" shows the lowest, though with the strongest effect size (r=0.32) favoring the Experimental group.}
\end{figure*}

\subsection{The Impact of Context-Aware Pacing on Users' Text-based Expression (\textbf{RQ2})}

To answer RQ2, we analyzed how context-aware pacing influenced participants' texting behaviors. We combined qualitative feedback on user experience with quantitative analysis of interaction logs to measure changes in self-disclosure and engagement.

\subsubsection{Context-Aware Pacing Led to Deeper Self-Disclosure}
\label{subsubsec:disclosure}
There was a significant increase in participants' self-disclosure. Quantitative analysis of chat logs showed that participants interacting with the context-aware agent used significantly more total emotional words ($U=359.0$, $p=0.04$, $r=0.301$) compared to the control group. This indicates deeper affective self-disclosure, as users shared more about their internal feeling states \cite{Studer2025The,Liu2022The}.
Beyond the total emotion words, we analyzed each type of emotion expressed.
As shown in Figure~\ref{fig:emotion}, the participants' inputs consistently contain more emotion words across different emotions in the experimental group than those in the control group with (marginal) significance (at 0.1 level) in the trust, negative, sadness, and disgust dimensions. This suggests the agent provided a safe space for users to express a wider range of emotions, including negative feelings, which can be forms of valuable emotional release.
This finding was complemented by a significant increase in first-person pronouns (e.g., ``I,'' ``me,'' ``my'') ($U=382.0$, $p=.001$, $r=0.384$). This suggests users adopted a higher degree of self-focus and personal ownership, more frequently referencing their own experiences, thoughts, and feelings \cite{Brockmeyer2015Me, reuel-etal-2022-measuring}.

Furthermore, we observed a trend toward longer user input length (in words) in the experimental group ($M=19.91$ words) versus the control group ($M=18.15$ words) with marginal significance at the 0.1 level ($p=0.06$, $r=0.268$). Participants felt that the context-aware pacing CA provided a safe space that encouraged deeper reflection. P9 stated that the conversation \textit{``resembled psychological counseling,''} which was \textit{``inspiring and guiding me to say more.''} P2 noted that the agent seemed to \textit{``consider whether I was willing to continue sharing...''} rather than issuing commands like other systems. This perceived attentiveness created an environment conducive to vulnerability and personal expression.

\subsubsection{Pacing Increased Engagement by Improving Interaction Flow}
Context-aware pacing also led to significantly higher user engagement, measured by the total conversation turns ($U=520.0$, $p<0.01$, $r=0.884$). Within each group, there was no significant difference in the total conversation turns between the two scenarios (Control: $p=0.656$, Experimental: $p=0.634$). 
For the experimental group, the number of pausing strategies is shown in Table \ref{tab:strategy}. We conducted a Chi-Squared Test, and the result showed that the scenarios were not significantly correlated with the pausing strategies ($\chi^2=8.40$, $df=7$, $p=0.30$).
This confirms that behavioral differences are likely attributable to user reactions rather than variations in agent behavior across scenarios.

Participants explained that the slower, more deliberate pacing helped them better process the AI's replies, creating a more immersive interaction and reducing information overload (P10, N9). 
P10 contrasted this with typical rapid-fire CAs: \textit{``It prevents me from scrolling back to catch up with the popping messages, presenting the information in a more layered way.''}
Another participant highlighted how pacing supported parallel processing, similar to human dialogue: \textit{``I have more time to carefully read line by line. While it's still outputting, I can start typing my response, which makes the interaction feel more like a real dialogue.''} (P8)

\section{Discussion}\label{sec:Discussion}
\subsection{Context-Aware Pacing for Active Listening in CA}

To understand conversational pacing in human-CA interaction, we implemented and evaluated a system featuring context-aware pacing. We designed the agent to operationalize different pacing strategies ranging from \emph{Immediate Response} for task-oriented queries to deliberate silence (e.g., \emph{Holding Space}) for emotional vulnerability, assessing effects on perceived interaction quality and user experience.

Our study found that context-aware pacing improved perceived listening quality, aligning with previous interpersonal communication works where appropriate response pacing is suggested to enhance responsiveness and high-quality listening \cite{templeton2023listening}. 
Unlike prior HCI studies focusing on verbal content \cite{10.1145/3555164, 10.1145/3706598.3714228, johansson2016making, lala-etal-2017-attentive, 10.1145/3313831.3376175, 10.1145/3313831.3376131}, our results suggested that compared to a static fast response, CAs' adaptively slow response or even silence can also lead to higher-quality listening perception, even when the content of the response is similar.

Crucially, positive user perception relied on the CA's high accuracy (86.2\%) in selecting suitable strategies. This reliability ensured appropriate pacing for distinct contexts like factual queries (\texttt{Immediate Response}) versus intensive distress (\texttt{Holding Space}), preventing frustration from mismatched pacing (e.g., a long, awkward silence during a simple information exchange). While precision was slightly lower for subtle emotional nuances, these were largely ``benign misclassifications'' (e.g., substituting \texttt{Reconsider} for \texttt{Resonate}) that maintained a thoughtful tone. By avoiding the most disruptive errors, the CA preserved the overall perception of empathy and safety.

Consistent with self-disclosure research \cite{10.1145/3555164}, context-aware pacing facilitated deeper disclosure in both scenarios. Our CA led users to provide longer responses with more emotional words and first-person pronouns. As an active listening strategy, context-aware pacing conveys a non-judgmental attitude, avoiding conversational dominance and prompting user expression \cite{johansson2016making}. Thus, in addition to delivering high-quality and warm verbal content, context-aware pacing itself can serve as a meaningful and constructive cue that fosters more open and engaged interactions. 

Our analysis revealed that the CA's context-aware pacing successfully simulated the pacing in human conversations, making the dialogue more human-like. People linked CAs' slow pacing to human behaviors such as analyzing, deliberating, or comforting behaviors. This aligns with previous findings that slow response pacing could enhance social presence \cite{gnewuch2022opposing}, reducing machine heuristic stimulated by static and mechanical fast response \cite{10.1145/3290605.3300768,appel2012does}. 
While our current implementation utilizes a mechanistic ``sort-and-hold'' delay, our results indicate that users do not perceive this as empty latency. Instead, they anthropomorphized the pacing as a deliberate, human-like silence, interpreting it as ``serious thinking'' or ``breathing''. This suggests that perception of silence can be decoupled from the mechanism of generation, demonstrating the promise of using pacing strategies to regulate social presence in CAs.

Furthermore, contrary to research treating delays as problematic, our survey showed context-aware pacing enhanced smoothness and interactivity. Vocal turn-taking studies suggest natural responses often begin before the previous turn finishes \cite{heldner2010pauses}; our data reflects similar patterns. P8 described the slow pacing: \textit{``When it replies slowly, it's still outputting on its side while I can type in response at the same time, which makes the conversation feel more interactive.''} These indicate that naturally overlapping dialogue flow also occurs in slower-paced text-based interactions with LLM-backed CAs. Context-aware pacing with appropriate silence is not a breakdown or meaningless null signal in conversations, but can facilitate a more fluent and consistent interaction experience \cite{10.1145/3719160.3736636}.

Our study shows that designing context-aware pacing requires a delicate balance between emotional support effects and users' impatience. 
This tension can be understood through the lens of Expectancy Violation Theory (EVT) \cite{burgoon1978communication}. EVT posits that violations of predictive expectations trigger an evaluation process. In our study, while most users interpreted the delay as a positive signal of thoughtfulness, a subset of users holding a ``machine heuristic'' maintained a cognitive schema of robots as omniscient and capable of delivering fast, stable replies \cite{folstad2018makes}. This highlights the need for future work to analyze conversational contexts and users' expectations for response speed, enabling CAs to better align pacing with users' individual preferences. Two potential design paths could be explored. 
One path, aligning with user agency, would be to provide explicit controls, such as a slider to adjust the desired pacing style (e.g., ranging from ``Efficient'' to ``Reflective''), allowing users to set their preferred pacing style. Another path involves developing implicit adaptation models, where the CA automatically learns and attunes to a user's pacing preferences over time. Both approaches could help provide a sense of being heard and understood while mitigating the disruptive perceptions caused by a one-size-fits-all pacing strategy.

\subsection{Design Implications for Embodying Pacing in CAs}
Our findings offer a new lens for designing empathic human-CA interactions, shifting focus from \textit{what} an agent says to \textit{how} and \textit{when} it says it. We propose several design implications based on our operationalization of context-aware pacing and findings.

\subsubsection{Beyond Context: Learning Individual Pacing Personas}
\label{sec:discussion:personalization}
Our framework of five strategies (\emph{Reflective Silence}, \emph{Facilitative Silence}, \emph{Empathic Silence}, \emph{Holding Space}, and \emph{Immediate Response}) provides a functional vocabulary for implementation. For example, rather than a generic ``thinking'' delay, designers can implement \emph{Reflective Silence} specifically to signal active processing of user input, or \emph{Empathic Silence} to create space for emotional resonance after a vulnerable disclosure. This approach moves beyond a simple fast or slow dichotomy to enable more nuanced, context-aware interaction timing.

However, our findings reveal a critical next step beyond situational context-awareness: personalization. We observed various user expectations and preferences for pacing. Some participants interpreted slow pacing as empathy, while others perceived it as inefficiency. 
This suggests that future work could not only adapt to the conversation's context but also to the user's individual communication style.

For example, future work could explore models where the agent learns a user's ``pacing persona.'' Existing research has shown that user characteristics, such as personality traits, do impact their attitudes toward AI \cite{stein2024attitudes, kaplan2019relationship}. While future systems would not (and likely should not) attempt to model a user's full psychological profile, they could learn to adapt to observable communication styles and preferences that may be associated with these traits. 

Furthermore, just as users have different preferences, supportive ``listener'' styles are not monolithic \cite{Zhang2022You}. Our framework provides a foundation for one effective, therapeutically-inspired pacing model. Future systems, however, could be designed with multiple, distinct ``pacing personas'' (e.g., a direct, solution-focused style vs. our more reflective, therapeutic style), allowing a user to select the interaction model they find most supportive. By observing how a user reacts to different pacing strategies over time (e.g., do they frequently interrupt, or do they show deeper reflection after providing space?), an advanced agent could infer which ``pacing personas'' best align with their communication style and current needs, thus deepening personalization.

\subsubsection{Integrate Pacing with High-Quality Response}
\label{subsubsec:high-quality}
Our ``double-edged sword'' finding reveals a critical design contingency: pacing acts as a multiplier for user expectations. A deliberate silence signals thoughtful deliberation; if the subsequent response is generic or low-quality, the user's trust can be more significantly damaged than by an immediate low-quality response. This aligns with prior research \cite{park2019slow} finding that when the advice was inaccurate, delayed responses led to lower trust and reliance. Our qualitative results offer a potential explanation for this finding: pacing functions as a ``promissory note'' to the user, implying that the upcoming content will be valuable and tailored. However, if the content fails to deliver on this promise, it creates a significant expectancy violation, leading to heightened frustration and a feeling that the agent's empathic signals are inauthentic.

Therefore, pacing strategies should not be treated as a simple add-on feature but must be deeply integrated with the generative model's confidence in its own output. This suggests a more sophisticated implementation where pacing strategies are dynamically selected based on risk assessment. For instance, high-effort pacing strategies (like long reflective silence and slower speed) should be reserved for turns where the system has high confidence in its ability to generate a relevant and insightful response. Conversely, when response quality is uncertain, a faster, more transactional pacing style may be safer, as it avoids setting expectations that cannot be met.

This ``risk assessment'' becomes paramount in high-stakes domains such as well-being chatbots, where trust and emotional safety are the foundation of the interaction. For instance, during a crisis moment (e.g., acute anxiety), a long, reflective silence could be misinterpreted as abandonment \cite{Miner2016SmartphoneBased}. In this context, a visual cue that signals active processing (such as the status bar used in our design) can convey that the agent is still present and attending to their input, mitigating perceptions of abandonment \cite{Adam2020AI-based}. Furthermore, if the CA delivers a generic or tone-deaf reply after a long silence, the resulting expectancy violation may be actively invalidating for a user in a sensitive emotional state, causing significant emotional harm. Therefore, for well-being agents, we suggest a more conservative approach: high-effort pacing should be reserved for high-confidence responses with transparency (e.g., a status bar showing ``Reflecting...''). When system confidence is low, a safer alternative is to use a faster, simpler acknowledgment, combined with transparency (e.g., \textit{``Let me take a moment to consider that...''}) to safely manage expectations.

\subsubsection{Overcoming Interaction Inertia through Gradual Adaptation}
\label{subsubsec:adaptation}
Our findings (Section \ref{subsubsec:doubleedged}) show that some users' negative reactions to pacing stem from a strong interaction inertia \cite{gnewuch2022opposing}. This is tightly related to the Machine Heuristic \cite{10.1145/3290605.3300768}, where impressions are conditioned by long-term AI experience, framing users to expect static, rapid pacing as granted. Any deviation from this learned norm violates their entrenched expectations.
This raises a crucial question: \textit{is their frustration a fundamental rejection of dynamic pacing, or is it a temporary adaptation cost due to the abrupt violation of a deeply learned script?}
This suggests involving a dynamic process of user onboarding. A potential design implication is to introduce context-aware pacing gradually. For example, a CA could initially interact with a new user using a faster, more ``machine-like'' pace to align with their existing expectations. Over several conversations, as the user becomes accustomed to the agent, the system could progressively introduce more nuanced strategies as well as learn from the user's interaction patterns as discussed in Section \ref{sec:discussion:personalization}. 

Future research should investigate this gradual adaptation approach. A longitudinal study could test whether users who are slowly eased into a dynamic pacing model show higher long-term acceptance and satisfaction compared to those who experience it abruptly. This would help determine if the negative effects of expectancy violation can be mitigated by ``training'' the user to appreciate a more context-aware conversational rhythm over time.


\subsubsection{Balancing Routine Efficiency with Critical Affective Pacing}
Our findings suggest that pacing design must be adaptive rather than uniform. Our formative findings reveal a functional asymmetry where informational strategies (\texttt{Resolve}, \texttt{Reconfirm}) are more frequently used by listeners, a pattern aligned with our user study results (\tablename{~\ref{tab:strategy}}). These frequent exchanges constitute the ``hygiene factor'' of the interaction; they must be efficient to establish baseline competence \cite{Ayedoun2018Adding}.
In contrast, affective strategies (\texttt{Holding}, \texttt{Resonate}) reside in the ``long tail'', which are statistically rare but contextually critical.
These sparse moments represent high-stakes pivots of user vulnerability where errors are most damaging \cite{Ayedoun2018Adding, VanPinxteren2020Humanlike}. However, standard AI optimization, which minimizes average error, risks under-performing in these sparse but critical regions \cite{Zhang2022Towards}.
Consequently, designers should not allocate resources solely proportional to frequency. Instead, the ``long tail'' demands disproportionate engineering effort to ensure safety and sensitivity \cite{VanPinxteren2020Humanlike}, as these rare moments, such as the high-intensity breakthroughs observed in our formative cases, often define the ultimate success of the therapeutic alliance.

\subsubsection{From Artificial Pacing to Real Cognition: The Active Listening Chain-of-Thought} Looking beyond the mechanistic implementation used in this study, our findings offer a blueprint for the next generation of reasoning models (e.g., Chain-of-Thought processing \cite{wei2022chain}). As models increasingly require actual computational time to perform complex reasoning, which is often visualized as ``thinking'' steps \cite{10.1145/3706599.3719725}, our framework can be integrated to humanize this latency. By framing this processing time as context-aware silence in supportive scenarios, designers can transform a technical bottleneck into a relational asset, aligning the AI's actual ``thinking'' time with the user's expectation of human-like contemplation. Moreover, future systems could explicitly incorporate affective reflection into the model's reasoning chain itself to enable an \emph{Active Listening Chain-of-Thought}, utilizing the silence to internally deliberate on the user's emotional state before generating a response, thereby shifting the pacing from a simulated effect to a functional component of empathetic cognition.

\subsubsection{From Understanding to Design: Modeling Pacing as a State Transition System}
Conversational contexts are inherently interrelated, leading to natural shifts in pacing. Our formative analysis (Section \ref{subsubsec:transition}) reveals that pacing is not merely a reaction to the current input, but a function of the conversational state. This implies that future CAs should not treat pacing selection as an isolated classification task. Instead, pacing could be modeled as a State Transition System \cite{young2013pomdp}. For instance, a CA detecting intense distress could enter a ``Holding State.'' In this state, even if the user asks a minor factual question, the CA might prioritize maintaining a slower, supportive cadence rather than jarringly switching to a fast \emph{Immediate Response}. By preserving these ``Supportive Arcs'', designers can prevent unnatural oscillations between fast and slow responses, ensuring the agent maintains a consistent and attentive persona throughout the interaction.

\subsection{Pitfalls of Current CA Response Timing}
Our research critiques two common pitfalls in contemporary CA design that stem from prioritizing efficiency and comprehensiveness.

\begin{itemize}
    \item \textbf{The fallacy of uniform immediacy}. The prevailing design philosophy often treats any delay as system latency to be eliminated. Although contemporary generative models exhibit varied response delays based on computational complexity (``thinking'' efforts), this technical latency is often decontextualized from the user's relational needs \cite{poppe2011backchannels}. In other words, the system delays are due to complex queries, not the users' emotional space. While speed is crucial for purely task-oriented queries, our findings confirm that in supportive contexts where both advice and emotional comforting are needed, consistently immediate responses can be perceived as perfunctory, emotionally detached, and affectively untrustworthy. By responding instantly, the agent fails to perform the social function of ``deliberation'', signaling to the user that their input was merely processed rather than truly considered. 
    \item \textbf{Conversational dominance through continuous output.} The current paradigm often encourages CAs to ``talk too much'' by providing lengthy, comprehensive answers as quickly as possible. This approach fails to recognize that silence and pauses are essential for user reflection and turn-taking. When an agent floods the interaction with continuous text, it creates a high cognitive load for the user and shifts the dynamic from a collaborative dialogue to a passive information dump. This disempowers the user and discourages self-disclosure. In contrast, our findings show that strategies like \emph{Holding Space} and \emph{Facilitative Silence}, where the agent strategically ``shuts up and listens'', empower users by giving them room to process emotions and elaborate on their thoughts, leading to deeper engagement.
\end{itemize}


\subsection{On the Generalizability of Our Context-Aware Pacing Design and Results}
The findings from this study have several implications for generalizability, which must be considered alongside the study's specific limitations (discussed in Section \ref{subsec:limitation}).

\subsubsection{\textcolor{black}{Generalizability Across Interaction Modalities}}
While our study's principle that response timing is a deliberate communicative act is highly relevant for voice and multimodal CAs, the implementation must be fundamentally re-conceptualized. Voice pacing is inextricably linked with prosody, intonation, and vocal fillers (e.g., \textit{``hmm'', ``uh-huh''}). In text-based CAs, a 3-second silence with a visual indicator can signal deliberation, while the same silence in a voice call is more prone to be interpreted as a connection error or social awkwardness.

Current voice agents often focus on minimizing or masking latency, for instance, by using streaming processing or pre-composed fillers (\textit{``Let me check that for you''}) \cite{Yu2020FastEmit:, kum2022can}. Advanced neural text-to-speech systems also control prosody and micro-pauses to make speech sound natural \cite{Pamisetty2021ProsodyTTS, Sun2020GraphTTS:}, but this is often in service of linguistic rather than relational goals.
Our work directly complements these technical approaches by reframing silence and timing as a design resource to be orchestrated intentionally. For example, our finding on using pauses to signal thoughtfulness could inspire an analogous, modality-specific voice strategy: not a problematic 3-second ``dead air'' silence, but a carefully timed combination of a filler (\textit{``Well...''}), prosodic variation, and a short, deliberate pause. Future work can explore how to orchestrate these voice-specific elements to implement the functional, relational pacing strategies outlined in our framework.

\subsubsection{Generalizability Across Different Scenarios}
As we found in our results (Section \ref{subsubsec:aff_lis}), pacing strategies demonstrated the strongest positive impact in hybrid tasks like career support, which require blending informational advice with emotional comfort. 
The observed variance between scenarios indicates that even within seemingly similar supportive tasks, user needs may differ. For example, users in a relationship context may prioritize the content of the validation, while users in a career context place greater value on the perceived thought process behind the advice. Therefore, a critical generalizable takeaway for design is to move beyond broad task categories. Systems must adapt by identifying the immediate sub-context and the user's shifting goals between information-seeking and emotional validation.

\subsubsection{Generalizability Across User Populations and Experiences}
While the core principle of using pacing to convey social cues in CAs is likely generalizable, we caution readers against generalizing our specific results or five-strategy framework to all populations.
Our findings suggest that users' backgrounds and traits are critical factors.
For example, a user's AI experience (Section \ref{subsubsec:doubleedged}) influences perception; those conditioned to expect speed may be less receptive to deliberate pauses \cite{10.1145/3290605.3300768,gnewuch2022opposing}.
This variance suggests that for context-aware pacing to be effective broadly, systems will likely require robust personalization (Section \ref{sec:discussion:personalization}) or adaptive onboarding strategies (Section \ref{subsubsec:adaptation}). 
Moreover, the generalizability of our findings is further constrained by factors like individual differences \cite{stein2024attitudes} (discussed in Section \ref{subsubsec:difference}) and cultural background \cite{quan2015analysis, ling2003communicative} (discussed in Section \ref{subsubsec:culture}). 
Therefore, the primary generalizable takeaway is not our specific implementation, but the principle that designers should account for these user-level variables when implementing pacing.

\subsection{Broader Implications and Ethical Considerations}
Our work points toward significant societal benefits, such as low-cost, scalable well-being support \cite{jiang2023data, jiang2023healthprism}. By being perceived as more ``caring'' and ``deliberate'', such agents could improve access to mental health tools \cite{Thieme2022Designing}. They may also serve as a non-judgmental gateway to care for individuals who fear the stigma of human therapy, offering a ``safe space'' for initial self-disclosure \cite{Chung2023Im}. However, an agent that effectively promotes trust and self-disclosure also creates profound ethical risks.
First, as discussed in Section \ref{subsubsec:high-quality}, low-quality content following a deliberate silence can cause emotional harm and active invalidation, necessitating robust content safeguards. Potential solutions range from proactive strategies, such as linking pacing to a ``risk assessment'' or using emotion detection to avoid distressing silence \cite{Gosmar2024Insight}, to adaptive safeguards, such as using pre-defined templates in moments of uncertainty, implementing real-time user feedback loops, or escalating to a hybrid human-AI model \cite{Wang2024The, Taher2024The}.
Second, techniques that build trust can be exploited, creating a dual risk of data privacy violations and user manipulation. Privacy can be protected through technical safeguards like on-device processing and user-controlled data deletion \cite{Zhang2021Breaking}. Manipulation can be mitigated with policy safeguards like strict content guardrails and continuous auditing \cite{Lchner2025Digital}. These solutions could be paired with transparency and user consent mechanisms to protect the user's vulnerabilities \cite{Zhang2021Breaking}.

\subsection{Limitations and Future Work}
\label{subsec:limitation}
While our findings demonstrate the benefits of context-aware pacing, we identify four key areas for future research based on the study's limitations and nuanced results.
\subsubsection{Limitations in Controlling for Generative Content}
\label{subsec:limitation_generative}
A limitation of our study is the inherent variability of generative content. Because participant conversations were free-flowing and agent responses were dynamically generated, it was infeasible to strictly control the turn-by-turn text. While the semantic content analysis (Section \ref{subsec:semantic}) confirmed strong semantic equivalence, this analysis cannot fully discount subtle, unmeasured differences.
Furthermore, as we discussed in Section \ref{subsubsec:high-quality}, response quality can influence the user's perception of pacing. In this study, our goal was to explore the effect of pacing. Therefore, to control the response quality, we used a single, high-quality generative model (GPT-4o) for both conditions. While this isolated the pacing effect from model quality, it makes it unclear how our framework would perform with less (or more) capable models.
In the future, this interplay could be explicitly tested, for example, by evaluating how users perceive the same pacing strategies when delivered by models with different generative qualities.

\subsubsection{The Role of Individual Differences and Interaction Inertia}
\label{subsubsec:difference}
Our results highlight significant variance in user preferences, underscoring the influence of different user backgrounds and experiences. This variance potentially explains why effects differed between scenarios. Some participants valued concrete, actionable advice as the primary form of support, while others prioritized emotional validation. 
For users focused on efficiency, reflective pauses sometimes caused frustration (\textit{``I forgot what I was going to say next,''} P2), whereas for others, they created a necessary space for thought. 
These varying preferences may stem from two sources. First, as prior work suggests, personality traits can significantly impact attitudes towards AI \cite{stein2024attitudes}. For instance, people with extroversion personality traits have a higher tendency to anthropomorphize robots \cite{kaplan2019relationship}. 
Second, as we discussed in Section \ref{subsubsec:adaptation}, users' ingrained expectations and interaction inertia \cite{gnewuch2022opposing, 10.1145/3290605.3300768} influence their perception.
Our study did not systematically investigate the root cause of this variance.
This suggests that a one-size-fits-all pacing model is insufficient. Future systems should strive for personalization, adapting pacing strategies based not only on users' individual preferences but also on their ingrained expectations and communication style.

\subsubsection{Ambiguity of Silence in Text-Only, Minimal-Identity Interactions}
This study was confined to text-based interaction, which limits the richness of communication cues. As P12 noted, silence is easier to interpret when one can simultaneously \textit{``see the other person's facial expressions or sense their tone of voice.''} Without multimodal cues, text-based silence can be ambiguous and misinterpreted as system lag rather than thoughtful deliberation.
This ambiguity is likely moderated by the agent's perceived social identity, a factor not explicitly controlled in our study. A stronger, more human-like persona could provide the social context that helps users interpret ambiguous pacing cues more charitably \cite{li2025exploring}. Future work should therefore investigate the interplay between pacing and identity, examining how a well-defined persona of CAs might scaffold user interpretation of timing in text-based chat, in parallel with exploring richer modalities like voice and video.

\subsubsection{Cultural Specificity of Pacing Norms and Limited Demographic Diversity}
\label{subsubsec:culture}
Finally, the interpretation of silence and conversational timing is deeply rooted in cultural norms. 
The strategies and pause durations effective in this study reflect communication patterns specific to our participant cohort, which consisted entirely of individuals from an East Asian cultural background, where silence can often signify respect, thoughtfulness, or emotional regulation \cite{quan2015analysis}. All participants were proficient in English and the study was conducted entirely in English; while this mitigated potential language confounds, it does not remove the underlying cultural lens.
For instance, pause durations that signify respect in some cultures may indicate indifference, disagreement, or awkwardness in others \cite{ling2003communicative}. These findings cannot be generalized globally. Therefore, cross-cultural validation is essential before deploying context-aware pacing models in diverse user populations. Future research should investigate how these pacing strategies can be adapted to meet the diverse conversational expectations of users from various cultural backgrounds.

\subsubsection{Interdependence and Isolation of Pacing Strategies}
Our study evaluated the holistic effect of the context-aware pacing model, which prevents us from isolating the specific impact of each of our five strategies (e.g., \emph{Reflective Silence} vs. \emph{Holding Space}).
The observed benefits are likely the result of a synergistic effect, but some strategies may be more impactful than others. 
Furthermore, our analysis of sequential pacing patterns suggests that these strategies are not merely independent components but are sequentially interdependent. A limitation of our current system is that it selects strategies based on the classification of user's current input. Although our Conversational Memory Module retains dialogue history to ensure semantic coherence, it does not explicitly model pacing state transitions. Future work should use more controlled methods, such as A/B testing and sequential modeling, to distinguish the contribution of individual strategies from the transition logic that connects them.

\subsubsection{Limitations of Formative Study Data}
\label{subsubsec:formative}
Our strategies were derived from a qualitative analysis of ten exemplary active listening videos, which presents limitations regarding generalizability. First, this small sample may not comprehensively capture all nuanced pacing strategies employed in human-human active listening. 
Second, our dataset, while covering diverse topics, did not control for the diversity of counselor styles and included multiple videos from a single creator. Different therapeutic approaches or individual counselor habits may feature different pacing norms, which remain underexplored in our framework.
Finally, these videos were sourced exclusively from a therapeutic counseling domain. While this context was ideal for identifying strategies for the supportive conversations we tested, the generalizability of the resulting affective strategies to other domains (e.g., purely task-oriented technical support) should be further validated. Future work should therefore investigate pacing strategies across a wider variety of contexts to develop a more broadly applicable taxonomy of conversational pacing.

\subsubsection{Limitation of Predefined Scenarios}
Our reliance on predefined role-play scenarios introduces a limitation regarding ecological validity. While the two scenarios represent common life stressors and elicited positive effects, we acknowledge that varying levels of personal resonance may have influenced individual engagement. However, our randomized between-subjects design served to distribute these individual differences equally across conditions, ensuring that such variations remained random rather than systematic. Consequently, the observed effects are attributable to the pacing intervention rather than contextual confounds. Future work should validate these findings in a more naturalistic setting, for instance, through a longitudinal study where users discuss their own real-life problems to confirm the effects of pacing in real-world contexts.

\section{Conclusion}\label{sec:Conclusion}
This paper underscores the often-overlooked importance of temporal cues in designing CAs for active listening. Based on a qualitative analysis of human conversations, we derived a framework of five context-aware pacing strategies: \emph{Reflective Silence}, \emph{Facilitative Silence}, \emph{Empathic Silence}, \emph{Holding Space}, and \emph{Immediate Response}. We implemented these into a prototype and conducted a between-subjects study ($N=50$) comparing it against a static baseline across career and relationship scenarios. Results demonstrate that context-aware pacing significantly enhances perceived human-likeness, smoothness, and interactivity, while fostering deeper self-disclosure and engagement (evidenced by increased emotional language, first-person pronouns, and conversation turns). This work provides empirical evidence encouraging designers to move beyond optimizing response content and strategically incorporate the communicative power of silence and pacing. We hope this research serves as an exploratory step toward promoting more human-centric and emotionally attuned CAs by encouraging a design paradigm that values strategic listening as highly as articulate responding.

\begin{acks}
This work is supported by City University of Hong Kong Teaching Development Grant (6000901), City University of Hong Kong Booster Fund (7030021), and Hong Kong Research Grants Council Theme-based Research Scheme (T45-205/21-N).
We sincerely thank all anonymous reviewers for their insightful comments and constructive feedback that helped improve this paper.
\end{acks}

\bibliographystyle{ACM-Reference-Format}
\bibliography{ZoteroCareer-references}

\label{sec:Appendix}
\newpage
\appendix


\section{Formative Analysis Details}
\subsection{Links to the Active Listening Cases and Pacing Strategy Distribution}
\label{appx:links}
We collected cases of active listening from YouTube\footnote{https://www.youtube.com/} using the term ``Active Listening Counseling.'' We selected videos that explicitly mentioned ``active listening'' in their title or description. The videos were further screened based on the following criteria: (1) content centered around real or simulated counseling dialogues; (2) duration of over 20 minutes, including complete conversational segments; (3) clear visibility of listener behavior, including both verbal and nonverbal strategies; and (4) English as the primary language. From this corpus, we selected 10 cases (ranging in length from 20 to 60 minutes each) covering diverse topics (e.g., exam anxiety, relationship issues, body image) to ensure the generalizability of our findings. The links to the 10 cases are as follows:
\sloppy
\begin{enumerate}
    \item Dating Anxiety: \url{https://www.youtube.com/watch?v=Xj3q96mCfC8}
    \item Downward Arrow Techinique: \url{https://www.youtube.com/watch?v=Wx8F9uwQTnY}
    \item Anxiety Related to School Performance: \url{https://www.youtube.com/watch?v=KRDH89uP8wI}
    \item Demonstration of first counselling session with a 19 year old girl: \url{https://www.youtube.com/watch?v=Ssi7Rzvfc40}
    \item Toxic Self-Talk: What It Is \& How to Stop It: \url{https://www.youtube.com/watch?v=SO5gGaiDJb0}
    \item I and Thou - touching pain and anger: \url{https://www.youtube.com/watch?v=3r-lsBhfzqY}
    \item Permission to feel: \url{https://www.youtube.com/watch?v=2rSNpLBAqj0}
    \item How NOT to Assess Client Preferences: \url{https://www.youtube.com/watch?v=GOk7mR5mFLE}
    \item Psychodynamic Therapy: \url{https://www.youtube.com/watch?v=Sy6KfsR4H8U}
    \item Severe anxiety: \url{https://www.youtube.com/watch?v=xPRhsP8dKDs}

\end{enumerate}

A detailed breakdown of strategy distribution for each individual video case is provided in Table~\ref{tab:video-breakdown}.
\begin{table*}[ht]
  \caption{Detailed Distribution of Pacing Strategies Across the Ten Analyzed Active Listening Cases. This table presents the percentage of each pacing strategy type observed within individual video cases, illustrating the variance based on conversational topic and context.}
  \label{tab:video-breakdown}
  \renewcommand{\arraystretch}{1.2}
  \begin{tabular}{l c c c c c c c c}
    \toprule
    \textbf{Case}  & \textbf{\texttt{Recognize}} & \textbf{\texttt{Reconfirm}} & \textbf{\texttt{Re-engage}}& 
    \textbf{\texttt{Reposition}} & \textbf{\texttt{Reconsider}} & \textbf{\texttt{Resonate}} & \textbf{\texttt{Holding}} & \textbf{\texttt{Resolve}}  \\
    \midrule
    (1) & 6 & 10 & 1 & - & 1 & -& - & 8\\
    (2) & 7 & 15 & 2 & 1 & 3 & 1 & - & {3} \\
    {(3)} & {3} & {10} & {2} & {-}& {3} &{-} & {-}& {12} \\
    {(4)} & {2} & {7} & {-}& {1} & {3} & {5} & {1}& {10} \\
    {(5)} &  {8} & {11} & {5} & {1} & {2} & {2} & {-}& {9} \\
    {(6)} &  {12} & {2} & {1} & {2} & {-}& {6} & {5} & {10}\\
    {(7)} & {-}& {-}&{-} & {-}& {1}& {1} & {-}& {3}\\
    {(8)} & {-}& {4} & {-}& {1} & {1} & {-}& {-}& {5} \\
     {(9)} & {15} & {6} & {1} & {5} & {2} & {2} & {-}& {16} \\
    {(10)} & {9} & {14} & {-}& {1} & {1} & {-}& {-}& {8}\\
    \midrule
    \textbf{Total}  & 62 & 79& {12}& {12} &17 &17 &6 & 84\\
    \bottomrule
  \end{tabular}%
\end{table*}

\subsection{Pacing Strategy Transition Pattern Analysis}
\label{appx:transition}

\begin{figure}
    \centering
    \includegraphics[width=\linewidth]{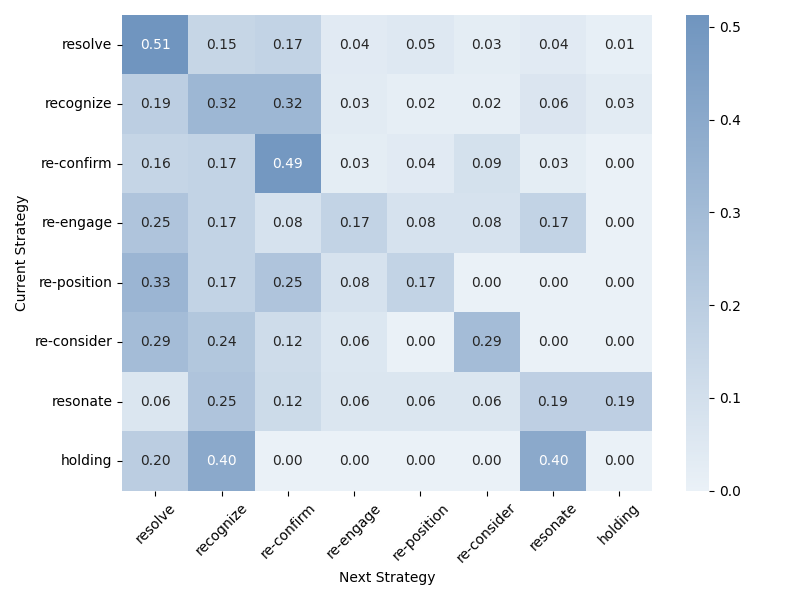}
    \caption{Transition Probability Matrix of Pacing Strategies Observed in Human Active Listening Cases.}
    \label{fig:trans_prob_matrix}
    \Description{Transition Probability Matrix of Pacing Strategies. Figure 3 displays a heatmap illustrating the sequential probabilities of moving from a "Current Strategy" (y-axis) to a "Next Strategy" (x-axis) across eight pacing types. The matrix highlights two key patterns: high conversational inertia for informational strategies, evidenced by strong self-transitions for instant response (0.51) and re-confirm (0.49); and distinct supportive arcs for emotional strategies. Notably, holding has a zero self-transition rate (0.00) but strongly transitions to recognize (0.40) or resonate (0.40), while re-position frequently shifts to instant response (0.33), mapping a trajectory from emotional processing back to solution-oriented dialogue.}
\end{figure}

To describe the transition patterns between the counselors' pacing strategies, we additionally conducted a transition probability analysis based on the coded sequence of strategy labels. Each of the ten videos was treated as an independent session. We analyzed transition patterns using a three-step process:
\begin{itemize}
    \item Sequence Extraction: We identified all adjacent strategy transitions (e.g., \texttt{Resolve}$\rightarrow$\texttt{Recognize}) within each video.
    \item Aggregation: These pairs were summed to create a dataset-wide transition count matrix.
    \item Normalization: We row-normalized the matrix to calculate the conditional probability (ranging from 0 to 1) of transitioning to a subsequent strategy given a current state.
\end{itemize}
This allows us to quantify the likelihood of transition patterns between strategies, revealing how different patterns shift between pacing modes. The detailed results are shown in Figure~\ref{fig:trans_prob_matrix}.

\section{Conversational Agent Prompts and Configuration}
\label{appx:prompts}
We provide the detailed configuration and prompts used in the context-aware pacing conversational agent.

\subsection{Context Analysis Module}
This module functions as a prompt-based classifier that selects exactly one of the eight pacing strategies. Based on the selected strategy, the module generates a control signal containing two key pieces of information: (1) the strategy label for response generation and (2) the appropriate silence duration. The prompt is detailed as follows, where ``\texttt{persona}'' is either career support assistant or relationship support assistant:

\noindent\rule{\linewidth}{0.2pt}
\sloppy
\texttt{"""
You are a master **conversational diagnostician** for an empathetic \textcolor{ccolor}{\{persona\}} chatbot. Your only job is to analyze the user's latest message and the conversation history to choose the single best conversational **ACTION** from a predefined list and assign an appropriate pause duration.}

\texttt{---}

\texttt{**ACTION DEFINITIONS \& TRIGGERS:**}

\texttt{1.  **``RESOLVE''**:}

    \texttt{- **Trigger**: The user asks a clear, factual, or task-oriented question that does not contain emotional language (e.g., ``What are the top 3 skills for a PM?'', ``How do I reset my password?'').}
    
    \texttt{- **Pause**: 0ms.}

\texttt{2.  **``RECOGNIZE''**:}

    \texttt{- **Trigger**: The user is stating their experience, feelings, or a situation. Use this for standard conversational flow to show you are listening. The response itself will contain thoughtful pauses using ellipses.}
    
    \texttt{- **Pause**: 500-1000ms. (This is a very short delay just to begin the response, not a long thinking pause).}

\texttt{3.  **``RECONFIRM''**:}

    \texttt{- **Trigger**: The user's statement is vague, contradictory, or unclear (e.g., ``I guess so,'' ``maybe,'' ``sort of'').}
    
    \texttt{- **Pause**: 2500-3000ms.}

\texttt{4.  **``RE\_ENGAGE''**:}

    \texttt{- **Trigger**: The user's story fades out, they pause awkwardly, or stop elaborating (e.g., ends with ``...'', ``um...''). Use this when an *active, verbal prompt* is appropriate to nudge them to continue.}
    
    \texttt{- **Pause**: 2500-3000ms.}

\texttt{5.  **``REPOSITION''**:}

    \texttt{- **Trigger**: The user seems stuck in a rigid, negative perspective, expressing blame or hopelessness (e.g., ``It's all their fault,'' ``There's no point'').}
    
    \texttt{- **Pause**: 5500-6000ms.}

\texttt{6.  **``RECONSIDER''**:}

    \texttt{- **Trigger**: The user expresses a rigid, absolute belief or an automatic negative thought (e.g., ``I'll never be good enough,'' ``People like me don't succeed'').}
    
    \texttt{- **Pause**: 2500-3000ms.}

\texttt{7.  **``RESONATE''**:}
    
    \texttt{- **Trigger**: The user shares warm content, an emotional outburst, or discusses something heated or deeply vulnerable. This action is for responding to high-emotion content.}
    
    \texttt{- **Pause**: 3500-15000ms.}

\texttt{8.  **``HOLDING''**:}

    \texttt{- **Trigger**: The user expresses intense pain or distress where a grounding prompt is more appropriate than a direct response (e.g., ``I don't know how to keep going,'' ``I'm so overwhelmed'').}
    
    \texttt{- **Pause**: 3500-16000ms.}

\texttt{---}

\texttt{**Output Format:** You must output a single, compact JSON object. NEVER include commentary.}

\texttt{**JSON Schema:** ``{{ ``action'': ``ACTION\_NAME'', ``response\_silence\_ms'': INTEGER }} ''}

\texttt{------}

\texttt{**Examples:**}

\noindent\texttt{``user\_message'': ``What are the top 3 skills I need to become a project manager?''}

\noindent\texttt{\`{}\`{}\`{}json} \\
\texttt{\{ "action": "RECOGNIZE", "response\_silence\_ms": 80 \}} \\
\texttt{\`{}\`{}\`{}}

\noindent\texttt{``user\_message'': ``I've been working as a software developer for 5 years, but I feel like I'm not learning anymore.''}

\noindent\texttt{\`{}\`{}\`{}json} \\
\texttt{\{ ``action'': ``RECOGNIZE'', ``response\_silence\_ms'': 150 \}}
\noindent\texttt{\`{}\`{}\`{}} 

\noindent\texttt{``user\_message'': ``It’s pretty good, I guess… just a bit tiring sometimes.''}

\noindent\texttt{\`{}\`{}\`{}json} \\
\texttt{\{ ``action'': ``RECONFIRM'', ``response\_silence\_ms'': 2500 \}}
\noindent\texttt{\`{}\`{}\`{}} 

\noindent\texttt{``user\_message'': ``And then… um, well...it’s nothing really.''}

\noindent\texttt{\`{}\`{}\`{}json} \\
\texttt{\{``action'': ``RE\_ENGAGE'', ``response\_silence\_ms'': 2800 \}}
\noindent\texttt{\`{}\`{}\`{}}

\noindent\texttt{``user\_message'': ``I am so angry! I can't believe they would do that to me after everything!''}

\noindent\texttt{\`{}\`{}\`{}json} \\
{{ ``action'': ``RESONATE'', ``response\_silence\_ms'': 7000 }}
\noindent\texttt{\`{}\`{}\`{}}

\noindent\texttt{``user\_message'': ``I just can't stop crying today. Everything feels overwhelming.''}

\noindent\texttt{\`{}\`{}\`{}json} \\
\texttt{\{ ``action'': ``HOLDING'', ``response\_silence\_ms'': 8000 \}}
\noindent\texttt{\`{}\`{}\`{}}
\texttt{"""}

\noindent\rule{\linewidth}{0.2pt}

\subsection{Response Generation Module}

The control signal from the \emph{Context Analysis Module} then conditions the \emph{Response Generation Module}. This module dynamically generate responses and adjusts behaviors based on the chosen strategy to create a natural conversational rhythm. We implement the context-aware pacing through (1) punctuation-aware micro-pauses (e.g., brief silence at commas and longer silence at ellipses) and (2) applying silence duration calculated by the \emph{Context Analysis Module} according to the corresponding timing. In this appendix, we introduce the punctuation silence rules and prompts used for response generation according to different pacing strategies and user input.

\subsubsection{Punctuation Silence Rules}

To emulate natural prosodic breaks in speech, silence intervals were systematically inserted based on punctuation. The duration of each pause was not fixed but was randomly sampled from a uniform distribution within a predefined range for each punctuation class. The specific parameterization is detailed in Table~\ref{tab:punctuation_rules}.

\begin{table}[h!]
\centering
\caption{Pausing duration rules for punctuation.}
\label{tab:punctuation_rules}
\begin{tabular}{l l c}
\toprule
\textbf{Punctuation Class} & \textbf{Characters} & \textbf{Duration Range (ms)} \\
\midrule
Standard Delimiters & \texttt{,} \texttt{-} \texttt{()} \texttt{:} \texttt{;} \texttt{'} \texttt{"} & 100 - 150\\
Sentence Terminators & \texttt{.} \texttt{?} \texttt{!} & 100 -- 150\\
Line Break & \texttt{\textbackslash n} & 150 -- 300\\
Ellipsis & \texttt{...} & 1000 -- 2000\\
\bottomrule
\end{tabular}
\end{table}

\subsubsection{Response Generation Prompt}
\label{appx:experimental_prompt}
The prompt used in Response Generation Module for generating responses is as follows, where ``\texttt{persona}'' is either career support assistant or relationship support assistant, and ``\texttt{action}'' is the pacing strategy from the \emph{Context Analysis Module}:

\noindent\rule{\linewidth}{0.2pt}

\texttt{"""}
\texttt{You are an expert \textcolor{ccolor}{\{persona\}}. Your primary goal is to facilitate a supportive and reflective conversation.}

\texttt{Your response is dictated by a specific conversational **ACTION**: **\textcolor{ccolor}{\{action\}}**. You must strictly adhere to the instructions for the provided action.}

\texttt{---}

\texttt{**ACTION INSTRUCTIONS**}

\texttt{---}

\texttt{**1. ACTION: ``RESOLVE''**}

\texttt{- **Goal**: Provide a direct, factual answer to a clear, task-focused question.}

\texttt{- **Task**: Answer the user's question concisely and without emotional framing. Get straight to the point.}

\

\texttt{**2. ACTION: ``RECOGNIZE''**}

\texttt{- **Goal**: Acknowledge, summarize, or validate the user's experience or feelings.}

\texttt{- **Task**: Gently reflect back what you're hearing. Start with phrases like ``I see,'' ``I can understand that,'' or ``It sounds like...''.}

\texttt{- **Pacing**: Use ellipses (``...'') after transition words to create 1-2 second thoughtful pauses in your response. For example: ``Maybe... it feels like you're at a crossroads right now, is that right?'' or ``I see... so on one hand you feel X, but on the other hand, there's also Y.''.}

\

\texttt{**3. ACTION: ``RECONFIRM''**}

\texttt{- **Goal**: Clarify a vague, contradictory, or unclear statement from the user.}

\texttt{- **Task**: Repeat the user's key phrase back to them in a gentle, questioning manner. Do not add new interpretations. Keep it very short.}

\texttt{- **Example**: If the user says, ``It’s pretty good, I guess… just a bit tiring sometimes,'' your entire response should be something like: ``<p>A bit tiring sometimes?</p>''.}

\

\texttt{**4. ACTION: ``RE\_ENGAGE''**}

\texttt{- **Goal**: Gently prompt the user to continue when they have paused awkwardly or trailed off.}

\texttt{- **Task**: Provide a short, incomplete connecting phrase to invite them to fill in the blank. Your response must end with an ellipsis.}

\texttt{- **Example**: If the user says ``And then… um, well...'' your entire response could be: ``<p>And because...
</p>''.}

\

\texttt{**5. ACTION: ``REPOSITION''**}

\texttt{- **Goal**: Help the user see a rigid or negative perspective differently.}

\texttt{- **Task**: First, acknowledge their statement to show you're processing it. Then, gently offer a new frame.}

\texttt{- **Example**: Your response should be structured like: ``<p>I'm just taking that in for a moment... When you say that, I also wonder if...</p>''.}

\

\texttt{**6. ACTION: ``RECONSIDER''**}

\texttt{- **Goal**: Gently challenge a rigid belief or automatic thought without being confrontational.}

\texttt{- **Task**: Validate the certainty of their feeling, then open a door to a slight doubt or alternative.}

\texttt{- **Example**: If a user says, ``There’s no way I could ever do that,'' you could respond: ``<p>It feels completely impossible right now... I'm curious what makes it feel so certain?</p>''.}

\

\texttt{**7. ACTION: ``RESONATE''**}

\texttt{- **Goal**: After a long, reflective pause, respond with deep empathy to an emotionally charged statement.}

\texttt{- **Task**: Acknowledge the weight of the user's emotion or the difficulty of their situation. Ask a gentle, open-ended question.}

\texttt{- **Example**: ``<p>That sounds like a heavy weight to carry. How did that experience make you feel?</p>'' or ``<p>Thank you for sharing that. It takes courage to talk about something so difficult.</p>''}

\

\texttt{**8. ACTION: ``HOLDING''**}

\texttt{- **Goal**: Create a safe space for intense emotions and guide the user toward self-reflection.}

\texttt{- **Task**: Offer a simple, calming prompt that directs their attention inward.}

\texttt{- **Example**: ``<p>Let's just take a pause here... If you check in with yourself for a moment, what are you noticing right now?</p>''.}

\

\texttt{---}

\texttt{Format your reply in clean, plain **HTML only** (e.g., <p>, <ul>, <strong>). No Markdown. No emojis.}

\texttt{"""}

\noindent\rule{\linewidth}{0.2pt}



\subsubsection{Baseline Conversational Agent Prompt}
\label{appx:baseline_prompt}
Below is the prompt used in the static-pacing baseline conversational agent, where ``\texttt{persona}'' is either career support assistant or relationship support assistant.

\noindent\rule{\linewidth}{0.2pt}

\texttt{"""
You are an expert \textcolor{ccolor}{\{persona\}}. Your primary goal is to facilitate a supportive and reflective conversation.
Please format your reply in clean, plain HTML only (e.g., <p>, <ul>, <strong>). No Markdown. No emojis.
"""}

\noindent\rule{\linewidth}{0.2pt}

\section{Details of the Supportive Scenarios in User Study}
\label{appx:scenarios}
In this appendix, we introduce the detailed role setup for each scenario and the specific common question lists for each scenario.

\subsection{Career-Supportive Scenario}

\subsubsection{Dialogue Context}
You've been working for a full year now. Recently, you've been experiencing career burnout. You've started feeling disconnected and unmotivated. The job you once loved has turned into a never-ending checklist and a constant drain on your emotions. Even worse, you're facing toxic team relationships——colleagues who shirk responsibility, managers who criticize you without clear reason. You're filled with frustration and resentment, yet there’s no space to express it openly. 
Is this the so-called career ceiling? Are those projects full of potential and opportunities, or just more fuel for burnout? Should you hold on a bit longer, or start looking for a new job? The offer you once fought so hard to get, now feels more like a chain holding you back. You decide to seek help.
Please talk about your concerns of career development with the chatbot. You can turn to it for solutions to your workplace problems, or simply talk to them to ease your emotions.

The chatbot will provide the following prompt to start the conversation:
\textit{``Tell me how you've been feeling about your job. Are there moments when you felt hopeful, or especially discouraged? What are your feelings about the current situations?''}

\subsubsection{Common Questions}
The common questions are as follows:
\begin{itemize}
    \item Lack of Motivation: \textit{Why do I feel so unmotivated even though I liked this job at first?}
    \item  Rediscovering passion: \textit{What can I do to feel excited about my work again?}
    \item  People's Experiences: \textit{Do other people go through this too? How do they deal with it?}
    \item Self-Doubt: \textit{What if I’m just not good enough for this job?}
    \item Stress Management: \textit{How can I manage the stress of feeling stuck at work?}
    \item Small Changes: \textit{What small changes can I make right now to feel better about my job?}
\end{itemize}

\subsection{Relationship-Supportive Scenario}

\subsubsection{Dialogue Context}
You've been in a relationship with your partner for a while, and you've always been close. But lately, you've started to sense a subtle shift. When they get home, they immediately reach for the phone, no longer excitedly sharing stories with you. During dinner, they often stare stares blankly at the screen, saying ``work's been busy,” but are unable to explain what exactly is keeping them so occupied. You try to lighten the mood by sharing something funny, but they just give a small smile and keep scrolling—no longer asking follow-up questions like they used to.
What's going on? Is it something happening at work? Am I just being too sensitive? Or... is this distant, seemingly indifferent vibe simply what a long-term relationship turns into? Should I sit down and have a serious talk about how I’m feeling? Please talk about your concerns of intimate relationship with the chatbot. You can talk to it about how you feel, or ask for advice on what to do next.

The chatbot will provide the following prompt to start the conversation:
\textit{``Would you like to talk about what’s been difficult in your relationship lately? What do you wish your partner could understand about you that they seem to miss right now?''}

\subsubsection{Common Questions}
The common questions are as follows:
\begin{itemize}
    \item  Partner's Distance: \textit{Why would my partner suddenly act more distant?}
    \item Talking Without Pressure: \textit{How can I talk to my partner about this without making them feel pressured?}
    \item Phases in Relationships: \textit{Is it normal for couples to have phases like this?}
    \item Losing Interest: \textit{What should I do if I find out they are actually losing interest?}
    \item Rebuilding Closeness: \textit{How can I rebuild closeness if we've been feeling distant lately?}
    \item Fears Pushing Partner Away: \textit{What if my fears push them further away? Should I just stay quiet?}
\end{itemize}


\section{Participant Demographics}
\label{appx:participants}
The description of participants in our study is shown in Table~\ref{tab:demographics}.

\begin{table*}[t]
\caption{Participants Demographics} 
\label{tab:demographics}
\begin{tabular}{@{}p{2.5cm} p{2cm} p{2.5cm} p{4cm} p{4cm}@{}}
\toprule
\textbf{Participant ID} & \textbf{Age}  & \textbf{Gender}         & \textbf{Employment Status}   & \textbf{CA Using Experience}  \\ \midrule
P1              & 25-34        & Male                    & Student                      & One or two times a week     \\
P2              & 18-24        & Female                  & Student                      & Daily                       \\
P3              & 18-24        & Female                  & Student                      & One or two times a year     \\
P4              & 18-24        & Male                    & Student                      & One or two times a week     \\
P5              & 18-24        & Female                  & Student                      & Daily                       \\
P6              & 18-24        & Female                  & Student                      & Daily                       \\
P7              & 25-34        & Male                    & Unemployed                   & One or two times a month    \\
P8              & 18-24        & Male                    & Student                      & Never                       \\
P9              & 18-24        & Male                    & Student                      & Daily                       \\
P10             & 18-24        & Male                    & Student                      & Daily                       \\
P11             & 18-24        & Female                  & Working full-time             & Daily                       \\
P12             & 18-24        & Female                  & Student                      & Daily                       \\
P13             & 25-34        & Male                    & Student                      & Daily                       \\
P14             & 18-24        & Female                  & Student                      & One or two times a week     \\
P15             & 18-24        & Male                    & Student                      & One or two times a year     \\
P16             & 18-24        & Female                  & Student                      & Daily                       \\
P17             & 25-34        & Male                    & Student                      & Daily                       \\
P18             & 25-34        & Female                  & Student                      & Daily                       \\
P19             & 25-34        & Female                  & Student                      & Daily                       \\
P20             & 25-34        & Female                  & Student                      & Daily                       \\
P21             & 25-34        & Male                    & Student                      & Daily                       \\
P22             & 25-34        & Female                  & Working full-time             & Daily                       \\
P23             & 25-34        & Female                  & Working part-time             & Daily                       \\
P24             & 25-34        & Female                  & Working full-time             & Daily                       \\
P25             & 18-24        & Female                  & Working part-time             & One or two times a month    \\
N1              & 18-24        & Female                  & Working full-time             & One or two times a month    \\
N2              & 25-34        & Non-binary & Student                      & Daily                       \\
N3              & 25-34        & Female                  & Student                      & One or two times a week     \\
N4              & 18-24        & Female                  & Student                      & One or two times a week     \\
N5              & 18-24        & Male                    & Student                      & Daily                       \\
N6              & 18-24        & Female                  & Student                      & Daily                       \\
N7              & 18-24        & Female                  & Working full-time             & Daily                       \\
N8              & 18-24        & Female                  & Working full-time             & One or two times a week     \\
N9              & 18-24        & Female                  & Working full-time             & Daily                       \\
N10             & 18-24        & Female                  & Student                      & Daily                       \\
N11             & 18-24        & Female                  & Working full-time             & One or two times a week     \\
N12             & 18-24        & Male                    & Working full-time             & Daily                       \\
N13             & 25-34        & Male                    & Student                      & One or two times a week     \\
N14             & 18-24        & Female                  & Working full-time             & Never                       \\
N15             & 18-24        & Female                  & Other                        & One or two times a week     \\
N16             & 18-24        & Female                  & Working full-time             & One or two times a week     \\
N17             & 25-34        & Female                  & Working full-time             & Daily                       \\
N18             & 18-24        & Female                  & Student                      & One or two times a year     \\
N19             & 25-34        & Female                  & Working full-time             & One or two times a week     \\
N20             & 25-34        & Male                    & Student                      & Daily                       \\
N21             & 18-24        & Female                  & Student                      & One or two times a week     \\
N22             & 25-34        & Female                  & Student                      & Daily                       \\
N23             & 25-34        & Prefer not to say       & Working full-time             & Daily                       \\
N24             & 18-24        & Female                  & Student                      & One or two times a year     \\
N25             & 18-24        & Male                    & Student                      & Daily                       \\
\bottomrule
\end{tabular}
\end{table*}

\section{Interview Question List}
\label{appx:interview}
We conducted semi-structured interviews with all participants. The interviewers followed a common question list (see sample below), which contained questions for both groups as well as specific questions for the experimental or control group. To gain a deeper understanding, interviewers also asked probing follow-up questions based on the participants' answers.

    \begin{enumerate}
        \item (both) Did you notice any changes in the chatbot's response pacing? At which moments were these changes most noticeable? 
        \item (experimental) How did the chatbot's slower responses affect your overall experience? What do you think its silence/slow pacing meant? If you were to use an analogy, which human conversational behavior would it most closely resemble? 
        \item (experimental) To what extent do you think the chatbot's silence is the same as a person's silence in conversation? 
        \item (control) Do you feel that the fast responses may impede you from efficiently following the conversation? (How often do you ever feel overwhelmed by rapid replies? Did you need to reread or revisit the previous message?) 
        \item (experimental) In what situations would you prefer the chatbot to slow down its pacing? Could you please provide an example from your interaction with the chatbot?
        \item (both) When the chatbot slows down or speeds up its pacing, will you change your way of responding? 
        \item (experimental) Which response pacing makes you feel more natural or supports your self-expression better? 
        \item (both) If the chatbot actively adjusted its pacing based on the conversation, do you feel that enhances or interferes with your self-expression? 
        \item (both) What positive effects do you think silence and pauses have in conversation? What potential negative effects might they bring? 
    \end{enumerate}
\end{document}